\documentclass[conference,a4paper]{IEEEtran}
\usepackage[dvips]{graphicx,color}
\usepackage{latexsym}
\usepackage{amssymb}
\usepackage{array}

\begin{document}

%
%
\newcommand{\qed}{\hfill$\square$}
\newcommand{\suchthat}{\mbox{~s.t.~}}
\newcommand{\markov}{\leftrightarrow}
\newenvironment{pRoof}{%
 \noindent{\em Proof.\ }}{%
 \hspace*{\fill}\qed \\
 \vspace{2ex}}


\newcommand{\ket}[1]{| #1 \rangle}
\newcommand{\bra}[1]{\langle #1 |}
\newcommand{\bol}[1]{\mathbf{#1}}
\newcommand{\rom}[1]{\mathrm{#1}}
\newcommand{\san}[1]{\mathsf{#1}}
\newcommand{\mymid}{:~}
\newcommand{\argmax}{\mathop{\rm argmax}\limits}
\newcommand{\argmin}{\mathop{\rm argmin}\limits}

\newcommand{\Cls}{class NL}
\newcommand{\vSpa}{\vspace{0.3mm}}
\newcommand{\Prmt}{\zeta}
\newcommand{\pj}{\omega_n}

\newfont{\bg}{cmr10 scaled \magstep4}
\newcommand{\bigzerol}{\smash{\hbox{\bg 0}}}
\newcommand{\bigzerou}{\smash{\lower1.7ex\hbox{\bg 0}}}
\newcommand{\nbn}{\frac{1}{n}}
\newcommand{\ra}{\rightarrow}
\newcommand{\la}{\leftarrow}
\newcommand{\ldo}{\ldots}
\newcommand{\typi}{A_{\epsilon }^{n}}
\newcommand{\bx}{\hspace*{\fill}$\Box$}
\newcommand{\pa}{\vert}
\newcommand{\ignore}[1]{}

%
%
%
%
\newcommand{\bc}{\begin{center}}  %
\newcommand{\ec}{\end{center}}
\newcommand{\befi}{\begin{figure}[h]}  %
\newcommand{\enfi}{\end{figure}}
\newcommand{\bsb}{\begin{shadebox}\begin{center}}   %
\newcommand{\esb}{\end{center}\end{shadebox}}
\newcommand{\bs}{\begin{screen}}     %
\newcommand{\es}{\end{screen}}
\newcommand{\bib}{\begin{itembox}}   %
\newcommand{\eib}{\end{itembox}}
\newcommand{\bit}{\begin{itemize}}   %
\newcommand{\eit}{\end{itemize}}
\newcommand{\defeq}{\stackrel{\triangle}{=}}

\newcommand{\Qed}{\hbox{\rule[-2pt]{3pt}{6pt}}}
\newcommand{\beq}{\begin{equation}}
\newcommand{\eeq}{\end{equation}}
\newcommand{\beqa}{\begin{eqnarray}}
\newcommand{\eeqa}{\end{eqnarray}}
\newcommand{\beqno}{\begin{eqnarray*}}
\newcommand{\eeqno}{\end{eqnarray*}}
\newcommand{\ba}{\begin{array}}
\newcommand{\ea}{\end{array}}
\newcommand{\vc}[1]{\mbox{\boldmath $#1$}}
\newcommand{\lvc}[1]{\mbox{\scriptsize \boldmath $#1$}}
\newcommand{\svc}[1]{\mbox{\scriptsize\boldmath $#1$}}

\newcommand{\wh}{\widehat}
\newcommand{\wt}{\widetilde}
\newcommand{\ts}{\textstyle}
\newcommand{\ds}{\displaystyle}
\newcommand{\scs}{\scriptstyle}
\newcommand{\vep}{\varepsilon}
\newcommand{\rhp}{\rightharpoonup}
\newcommand{\cl}{\circ\!\!\!\!\!-}
\newcommand{\bcs}{\dot{\,}.\dot{\,}}
\newcommand{\eqv}{\Leftrightarrow}
\newcommand{\leqv}{\Longleftrightarrow}
\newtheorem{co}{Corollary} 
\newtheorem{lm}{Lemma} 
\newtheorem{Ex}{Example} 
\newtheorem{Th}{Theorem}
\newtheorem{df}{Definition} 
\newtheorem{pr}{Property} 
\newtheorem{pro}{Proposition} 
\newtheorem{rem}{Remark} 

\newcommand{\lcv}{convex } 

\newcommand{\hugel}{{\arraycolsep 0mm
                    \left\{\ba{l}{\,}\\{\,}\ea\right.\!\!}}
\newcommand{\Hugel}{{\arraycolsep 0mm
                    \left\{\ba{l}{\,}\\{\,}\\{\,}\ea\right.\!\!}}
\newcommand{\HUgel}{{\arraycolsep 0mm
                    \left\{\ba{l}{\,}\\{\,}\\{\,}\vspace{-1mm}
                    \\{\,}\ea\right.\!\!}}
\newcommand{\huger}{{\arraycolsep 0mm
                    \left.\ba{l}{\,}\\{\,}\ea\!\!\right\}}}

\newcommand{\Huger}{{\arraycolsep 0mm
                    \left.\ba{l}{\,}\\{\,}\\{\,}\ea\!\!\right\}}}

\newcommand{\HUger}{{\arraycolsep 0mm
                    \left.\ba{l}{\,}\\{\,}\\{\,}\vspace{-1mm}
                    \\{\,}\ea\!\!\right\}}}

\newcommand{\hugebl}{{\arraycolsep 0mm
                    \left[\ba{l}{\,}\\{\,}\ea\right.\!\!}}
\newcommand{\Hugebl}{{\arraycolsep 0mm
                    \left[\ba{l}{\,}\\{\,}\\{\,}\ea\right.\!\!}}
\newcommand{\HUgebl}{{\arraycolsep 0mm
                    \left[\ba{l}{\,}\\{\,}\\{\,}\vspace{-1mm}
                    \\{\,}\ea\right.\!\!}}
\newcommand{\hugebr}{{\arraycolsep 0mm
                    \left.\ba{l}{\,}\\{\,}\ea\!\!\right]}}
\newcommand{\Hugebr}{{\arraycolsep 0mm
                    \left.\ba{l}{\,}\\{\,}\\{\,}\ea\!\!\right]}}

\newcommand{\HugebrB}{{\arraycolsep 0mm
                    \left.\ba{l}{\,}\\{\,}\vspace*{-1mm}\\{\,}\ea\!\!\right]}}

\newcommand{\HUgebr}{{\arraycolsep 0mm
                    \left.\ba{l}{\,}\\{\,}\\{\,}\vspace{-1mm}
                    \\{\,}\ea\!\!\right]}}

\newcommand{\hugecl}{{\arraycolsep 0mm
                    \left(\ba{l}{\,}\\{\,}\ea\right.\!\!}}
\newcommand{\Hugecl}{{\arraycolsep 0mm
                    \left(\ba{l}{\,}\\{\,}\\{\,}\ea\right.\!\!}}
\newcommand{\hugecr}{{\arraycolsep 0mm
                    \left.\ba{l}{\,}\\{\,}\ea\!\!\right)}}
\newcommand{\Hugecr}{{\arraycolsep 0mm
                    \left.\ba{l}{\,}\\{\,}\\{\,}\ea\!\!\right)}}

\newcommand{\hugepl}{{\arraycolsep 0mm
                    \left|\ba{l}{\,}\\{\,}\ea\right.\!\!}}
\newcommand{\Hugepl}{{\arraycolsep 0mm
                    \left|\ba{l}{\,}\\{\,}\\{\,}\ea\right.\!\!}}
\newcommand{\hugepr}{{\arraycolsep 0mm
                    \left.\ba{l}{\,}\\{\,}\ea\!\!\right|}}
\newcommand{\Hugepr}{{\arraycolsep 0mm
                    \left.\ba{l}{\,}\\{\,}\\{\,}\ea\!\!\right|}}

\newcommand{\MEq}[1]{\stackrel{
{\rm (#1)}}{=}}

\newcommand{\MLeq}[1]{\stackrel{
{\rm (#1)}}{\leq}}

\newcommand{\ML}[1]{\stackrel{
{\rm (#1)}}{<}}

\newcommand{\MGeq}[1]{\stackrel{
{\rm (#1)}}{\geq}}

\newcommand{\MG}[1]{\stackrel{
{\rm (#1)}}{>}}

\newcommand{\MPreq}[1]{\stackrel{
{\rm (#1)}}{\preceq}}

\newcommand{\MSueq}[1]{\stackrel{
{\rm (#1)}}{\succeq}}

\newenvironment{jenumerate}
	{\begin{enumerate}\itemsep=-0.25em \parindent=1zw}{\end{enumerate}}
\newenvironment{jdescription}
	{\begin{description}\itemsep=-0.25em \parindent=1zw}{\end{description}}
\newenvironment{jitemize}
	{\begin{itemize}\itemsep=-0.25em \parindent=1zw}{\end{itemize}}
\renewcommand{\labelitemii}{$\cdot$}

\newcommand{\iro}[2]{{\color[named]{#1}#2\normalcolor}}
\newcommand{\irr}[1]{{\color[named]{Red}#1\normalcolor}}

\newcommand{\irg}[1]{{\color[named]{Green}#1\normalcolor}}
\newcommand{\irb}[1]{{\color[named]{Blue}#1\normalcolor}}
\newcommand{\irBl}[1]{{\color[named]{Black}#1\normalcolor}}
\newcommand{\irWh}[1]{{\color[named]{White}#1\normalcolor}}

\newcommand{\irY}[1]{{\color[named]{Yellow}#1\normalcolor}}
\newcommand{\irO}[1]{{\color[named]{Orange}#1\normalcolor}}
\newcommand{\irBr}[1]{{\color[named]{Purple}#1\normalcolor}}
\newcommand{\IrBr}[1]{{\color[named]{Purple}#1\normalcolor}}
\newcommand{\irBw}[1]{{\color[named]{Brown}#1\normalcolor}}
\newcommand{\irPk}[1]{{\color[named]{Magenta}#1\normalcolor}}
\newcommand{\irCb}[1]{{\color[named]{CadetBlue}#1\normalcolor}}

%
\newenvironment{indention}[1]{\par
\addtolength{\leftskip}{#1}\begingroup}{\endgroup\par}
%
\newcommand{\namelistlabel}[1]{\mbox{#1}\hfill} 
\newenvironment{namelist}[1]{%
\begin{list}{}
{\let\makelabel\namelistlabel
\settowidth{\labelwidth}{#1}
\setlength{\leftmargin}{1.1\labelwidth}}
}{%
\end{list}}
%
%
\newcommand{\bfig}{\begin{figure}[t]}
\newcommand{\efig}{\end{figure}}
\setcounter{page}{1}

\newtheorem{theorem}{Theorem}

\newcommand{\ep}{\mbox{\rm e}}

\newcommand{\Exp}{\exp
}
\newcommand{\idenc}{{\varphi}_n}
\newcommand{\resenc}{
{\varphi}_n}
\newcommand{\ID}{\mbox{\scriptsize ID}}
\newcommand{\TR}{\mbox{\scriptsize TR}}
\newcommand{\Av}{\mbox{\sf E}}

\newcommand{\Vl}{|}
\newcommand{\Ag}{(R,P_{X^n}|W^n)}
\newcommand{\Agv}[1]{({#1},P_{X^n}|W^n)}
\newcommand{\Avw}[1]{({#1}|W^n)}

\newcommand{\Jd}{X^nY^n}
\newcommand{\IdR}{r_n}
\newcommand{\Index}{{n,i}}
\newcommand{\cid}{C_{\mbox{\scriptsize ID}}}
\newcommand{\cida}{C_{\mbox{{\scriptsize ID,a}}}}
\newcommand{\One}{\rm (i)}
\newcommand{\Two}{\rm (ii)}
\newcommand{\Thr}{\rm (iii)}
\newcommand{\Fou}{\rm (iv)}
\newcommand{\Fiv}{\rm (v)}

\newcommand{\Dist}{\Delta}
\newcommand{\pANDq}{q||p}
\newcommand{\ptANDqt}{q_t||p_t}
\newcommand{\OMega}{\Omega^{(\alpha,\mu,\lambda)}(q^n,S_n |p_{XY})}
\newcommand{\OMeGa}{\Omega^{(\alpha,\mu,\lambda)}(q|p_{XY})}
\newcommand{\ARgRv}{(p^{(n)},q^{n}|p_{XY})}
\newcommand{\pNqN}{\empty}

\newcommand{\BiBArXiv}{
}
\newcommand{\ArXiv}{. }
\newcommand{\GorF}{given }
\newcommand{\GorFb}{ will be given in the next section.}
\newcommand{\Comment}{}
\arraycolsep 0.5mm
\date{}
%
\title{
Exponent Function for Source Coding with Side Information at 
the Decoder at Rates below the Rate Distortion Function 
}

\author{%
\IEEEauthorblockN{Yasutada Oohama}
\IEEEauthorblockA{
  University of Electro-Communications, Tokyo, Japan \\
  Email: oohama@uec.ac.jp} 
} 

\newcommand{\Empty}{
\author{%
Yasutada Oohama 
\thanks{
Y. Oohama is with 
University of Electro-Communications,
1-5-1 Chofugaoka Chofu-shi, Tokyo 182-8585, Japan.
}%
}
\markboth{
}
{
}

}
\maketitle

\begin{abstract}
We consider the rate distortion problem with side information at the decoder 
posed and investigated by Wyner and Ziv. The rate distortion function 
indicating the trade-off between the rate on the data compression and 
the quality of data obtained at the decoder was determined by Wyner and Ziv. 
In this paper, we study the error probability of decoding at rates below 
the rate distortion function. We evaluate the probability 
of decoding such that the estimation of source outputs by the 
decoder has a distortion not exceeding a prescribed distortion level. 
We prove that when the rate of the data compression is below 
the rate distortion function this probability goes to zero exponentially  
and derive an explicit lower bound of this exponent function. 
On the Wyner-Ziv source coding problem the strong converse 
coding theorem has not been established yet. We prove this 
as a simple corollary of our result.  
\end{abstract}

\newcommand{\Zapss}{
\begin{keywords} 
source coding with side information at the decoder, 
the rate distortion function, exponent function at rates 
below the rate distortion function, strong converse theorem 
\end{keywords}
}

\newcommand{\ByShun}[1]{{\color[named]{Black}#1\normalcolor}}

\newcommand{\Zap}{
\section{Introduction}

We consider the rate distortion problem with side information at the decoder 
posed and investigated by Wyner and Ziv. The rate distortion function 
indicating the trade-off between the rate on the data compression and 
the quality of data obtained at the decoder was determined by Wyner and Ziv. 
In this paper, we study the error probability of decoding at rates below 
the rate distortion function. We evaluate the probability 
of decoding such that the estimation of source outputs by the 
decoder has a distortion not exceeding a prescribed distortion level. 
We prove that when the rate on the data compression is below 
the rate distortion function this probability goes to zero exponentially  
and derive an explicit lower bound of this exponent function. 
On the Wyner-Ziv source coding problem the strong converse 
coding theorem has not been established yet. We prove this 
as a simple corollary of our result.  
}

\section{
Source coding with Side Information at the Decoder
}

Let ${\cal X}$ and ${\cal Y}$ be finite sets and 
$\left\{(X_{t},Y_{t})\right\}_{t=1}^{\infty}$ be a stationary 
discrete memoryless source. For each $t=1,2,\cdots$, the random 
pair $(X_{t},Y_{t})$ takes values in ${\cal X}\times {\cal Y}$, 
and has a probability distribution  
$$
p_{XY}=\left\{p_{XY}(x,y)\right\}_{(x,y)\in {\cal X} \times {\cal Y}}
$$ 
We write $n$ independent copies of 
$\left\{X_{t}\right\}_{t=1}^{\infty}$ and 
$\left\{Y_{t}\right\}_{t=1}^{\infty}$, 
respectively as
$$
X^{n}=X_1,X_2,\cdots,X_{n}
\mbox{ and }Y^{n}= Y_1,Y_2,\cdots,Y_{n}.
$$
We consider a communication system depicted in Fig. 1.
Data sequences $X^{n}$ 
is separately 
encoded to 
$\varphi^{(n)}(X^{n})$ and is sent to 
the information processing center.
At the center the decoder function $\psi^{(n)}$ observes 
$\varphi^{(n)}(X^{n})$ and $Y^{n}$ to output 
the estimation $Z^n
$ of ${X}^{n}$. The encoder 
function $\varphi^{(n)}$ is defined by
\beq
\ba{l}
\varphi^{(n)}:{\cal X}^{n}\to {\cal M}_n
=\left\{\,1,2,\cdots, M_n\,\right\},
\ea
\label{eqn:defen1} 
\eeq
where $\| \varphi^{(n)}\|$ $(=M_n)$ stands for the range 
of cardinality of $\varphi^{(n)}$. Let ${\cal Z}$ be a 
reproduction alphabet. The decoder function 
$\psi^{(n)}$ is defined by
\begin{equation}
\psi^{(n)}:{\cal M}_n \times {\cal Y}^n 
\,\to\,{\cal Z}^{n}.
\end{equation}
%
%
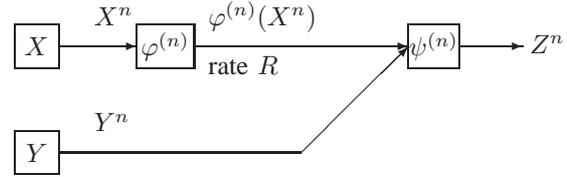
\begin{figure}[t]
\setlength{\unitlength}{0.94mm}

\begin{picture}(84,35)(4,4)
\put(10,25){\framebox(6,6){$X$}}
\put(10,10){\framebox(6,6){$Y$}}
\put(21,31){${X^n}$}
\put(16,28){\vector(1,0){11}}
\put(21,16){${Y^n}$}
\put(16,13){\line(1,0){11}}

\put(27,25){\framebox(8,6){$\varphi^{(n)}$}}
\put(37,31){$\varphi^{(n)}({X^n})$}
\put(37,24){rate $R$}
\put(27,13){\line(1,0){8}}
\put(50,28){\vector(1,0){15}}




\put(35,28){\line(1,0){15}}
\put(35,13){\line(1,0){15}}
\put(50,13){\vector(1,1){15}}


\put(65,25){\framebox(7,6){$\psi^{(n)}$}}

\put(72,28){\vector(1,0){9}}
\put(82,27){$Z^n$}

\end{picture}
\caption{
Wyner-Ziv source coding system.
}
\label{fig:Zigzag}
\noindent
\end{figure}
%
%
Let $d: {\cal X} \times {\cal Z}$ $\to [0,\infty)$ be 
an arbitrary 
distortion measure on ${\cal X} \times {\cal Z}$.
The distortion between $x^n\in {\cal X}^n$ 
and $z^n\in {\cal Z}^n$ is defined by
$$
d(x^n,z^n)\defeq \sum_{t=1}^n d(x_t,z_t). 
$$ 
The excess-distortion probability of decoding is 
\begin{equation}
{\rm P}_{\rm e}^{(n)}(\varphi^{(n)},\psi^{(n)};{\Dist})
=\Pr\left\{\frac{1}{n} d(X^n,Z^{n})\geq {\Dist} \right\}, 
\end{equation}
where $Z^{n}
=\psi^{(n)}(\varphi^{(n)}(X^{n}),Y^{n})$.
The average distortion 
$\Delta^{(n)}$ between $X^n$ and $Z^n$ is defined by 
$$
\Delta^{(n)}\defeq \frac{1}{n}{\rm E}\left[d(X^n,Z^n) \right]
\defeq \frac{1}{n}\sum_{t=1}^n {\rm E}d(X_t,Z_t).
$$
A pair $(R,\Dist)$ is $\varepsilon$-{\it achievable} for $p_{XY}$ 
if there exist a sequence of pairs 
$\{(\varphi^{(n)},$ $\psi^{(n)})\}_{n\geq 1}$
such that for any $\delta>0$ and any $n$ with 
$n\geq n_0$$=n_0(\varepsilon,\delta)$, 
\beqno
& &
\frac{1}{n}\log \| \varphi^{(n)}\| \leq R+\delta, 
\quad
{\rm P}_{\rm e}^{(n)}
(\varphi^{(n)},\psi^{(n)};{\Dist})\leq \varepsilon.
\eeqno
The rate distortion region ${\cal R}_{\rm WZ }(\varepsilon|p_{XY})$ 
is defined by 
\beqno
& &{\cal R}_{\rm WZ}(\varepsilon|p_{XY})
\\
&=&\left\{\,(R,\Dist):(R,\Dist)\,
\mbox{ is $\varepsilon$-achievable for }p_{XY}\,\right\}.
\eeqno
Furthermore set
$$
{\cal R}_{\rm WZ}(p_{XY})
\defeq \bigcap_{\varepsilon>0} {\cal R}_{\rm WZ}(\varepsilon|p_{XY}).
$$
On the other hand, we can define a rate distortion region 
based on the average distortion criterion, a formal definition 
of which is the following. A pair $(R,{\Dist})$ is {\it achievable} for $p_{XY}$ 
if there exist a sequence of pairs 
$\{(\varphi^{(n)},$ $\psi^{(n)})\}_{n\geq 1}$
such that for any $\delta>0$ and any $n$ with 
$n\geq n_0$$=n_0(\delta)$, 
\beqno
& &
\frac{1}{n}\log \| \varphi^{(n)}\| \leq R+\delta,\quad 
\Delta^{(n)}\leq {\Dist}+\delta.
\eeqno
The rate distortion region $\tilde{\cal R}_{\rm WZ }(p_{XY})$ is defined by 
\beqno
& &\tilde{\cal R}_{\rm WZ}(p_{XY})
\\
& \defeq&\left\{\,(R,\Dist):(R,\Dist)\,
\mbox{ is achievable for }p_{XY}\,\right\}.
\eeqno
We can show that the three rate distortion regions 
${\cal R}_{\rm WZ}(\varepsilon|$ $p_{XY})$, 
$\varepsilon \in(0,1)$, 
${\cal R}_{\rm WZ }(p_{XY})$, 
and $\tilde{\cal R}_{\rm WZ }(p_{XY})$ 
satisfy the following property. 
\begin{pr}\label{pr:pro0a}
$\quad$
\begin{itemize}
\item[a)] 
The regions 
${\cal R}_{\rm WZ }(\varepsilon|p_{XY})$, 
$\varepsilon \in(0,1)$, 
${\cal R}_{\rm WZ }(p_{XY})$, 
and $\tilde{\cal R}_{\rm WZ }(p_{XY})$ 
are closed convex sets of 
$\mathbb{R}_{+}^2$, where
\beqno
\mathbb{R}_{+}^2&\defeq &\{(R,\Dist): R\geq 0, \Dist\geq 0\}.
\eeqno

\item[b)] 
${\cal R}_{\rm WZ}(\varepsilon|p_{XY})$ has another 
form using $(n,\vep)$-rate distortion region, 
the definition of which is as follows.
We set 
\beqno
& &{\cal R}_{\rm WZ}(n,\varepsilon |p_{XY})
\\
&=&\{(R,\Delta): \mbox{ There exists }
(\varphi^{(n)},\psi^{(n)})
\mbox{ such that }
\\
& &\frac{1}{n}\log ||\varphi^{(n)}||\leq R, 
\quad {\rm P}_{\rm e}^{(n)}(\varphi^{(n)},\psi^{(n)};{\Dist})
\leq \varepsilon\},
\nonumber
\eeqno 
which is called the $(n,\vep)$-rate distortion region. 
Using ${\cal R}_{\rm WZ}(n,\varepsilon| p_{XY})$, 
${\cal R}_{\rm WZ}(\varepsilon|p_{XY})$ can be 
expressed as 
\beqno
{\cal R}_{\rm WZ}(\varepsilon | p_{XY})
&=&{\rm cl}\left(\bigcup_{m\geq 1}
\bigcap_{n \geq m}{\cal R}_{\rm WZ}(n,\varepsilon | p_{XY})
\right),
\eeqno
where ${\rm cl}(\cdot)$ stands for the closure operation.

%
\end{itemize}
\end{pr}

Proof of this property is \GorF in Appendix \ref{sub:ApdaAaaa}\ArXiv
\newcommand{\ApdaAaaa}{
\subsection{Properties of the Rate Distortion Regions}
\label{sub:ApdaAaaa}

In this appendix we prove Property \ref{pr:pro0a}.
Property \ref{pr:pro0a} part a) can easily be proved by 
the definitions of the rate distortion regions. 
We omit the proofs of this part. In the following 
argument we prove part b). 

{\it Proof of Property \ref{pr:pro0a} part b:} \ 
We set
\beqno
\underline{\cal R}_{\rm WZ}(m, \varepsilon| p_{XY})
&=&
\bigcap_{n \geq m}{\cal R}_{\rm WZ}(n,\varepsilon | p_{XY}).
\eeqno
By the definitions of 
$\underline{\cal R}_{\rm WZ}(m,\vep|p_{XY})$ and 
${\cal R}_{\rm WZ}(\vep|p_{XY})$, 
we have that 
$ \underline{\cal R}_{\rm WZ}(m,\varepsilon|p_{XY}) 
$ $\subseteq {\cal R}_{\rm WZ}(\varepsilon|p_{XY}) 
$
for $m\geq 1$. Hence we have that 
\beq
\bigcup_{m\geq 1} \underline{\cal R}_{\rm WZ}(m,\varepsilon|p_{XY}) 
\subseteq {\cal R}_{\rm WZ}(\varepsilon|p_{XY}). 
\label{eqn:aaDs}
\eeq
We next assume that $(R,\Dist)\in {\cal R}_{\rm WZ}(\varepsilon|p_{XY})$.
Set
$$
{\cal R}_{\rm WZ}^{(\delta)}(\varepsilon|p_{XY})
\defeq\{(R+\delta,\Dist): 
(R,\Dist)\in {\cal R}_{\rm WZ}(\varepsilon|p_{XY})
\}
$$
Then, by the definitions of 
${\cal R}_{\rm WZ}(n,\vep$ $| p_{XY})$ 
and ${\cal R}_{\rm WZ}($ $\vep|p_{XY})$, we have that 
for any $\delta>0$, there exists $n_0(\delta)$ 
such that for any $n\geq n_0(\delta)$, 
$$
(R+\delta,\Dist) \in {\cal R}_{\rm WZ}
(n,\varepsilon|p_{XY}),
$$
which implies that 
\beqa
& &{\cal R}^{(\delta)}_{\rm WZ}(\varepsilon|p_{XY})
\subseteq  \bigcup_{n\geq n_0(\delta)} 
{\cal R}_{\rm WZ}(n,\varepsilon|p_{XY})
\nonumber\\
&= & 
\underline{\cal R}_{\rm WZ}(n_0(\delta),\varepsilon|p_{XY})
\nonumber\\
&\subseteq & {\rm cl}\left(\bigcup_{m \geq 1} 
\underline{\cal R}_{\rm WZ}(m,\varepsilon|p_{XY})\right).
\label{eqn:aaDsX}
\eeqa
Here we assume that there exists a pair $(R,\Dist)$ 
belonging to ${\cal R}_{\rm WZ}(\vep | p_{XY})$ such that
\beq
(R,\Dist)\notin 
{\rm cl}\left(
\bigcup_{m \geq 1}
\underline{\cal R}_{\rm WZ}(m,\varepsilon|p_{XY})
\right).
\label{eqn:zdff}
\eeq
Since the set in the right hand side of (\ref{eqn:zdff}) 
is a closed set, we have 
\beq
(R+\delta,\Dist)\notin 
{\rm cl}\left(
\bigcup_{m \geq 1}
\underline{\cal R}_{\rm WZ}(m,\varepsilon|p_{XY})
\right)
\label{eqn:aas}
\eeq
for some small $\delta>0$. Note that $(R+\delta,\Dist)$ 
$\in {\cal R}^{(\delta)}_{\rm WZ}(\varepsilon|p_{XY})$.
Then (\ref{eqn:aas}) contradicts (\ref{eqn:aaDsX}). Thus we 
have 
\beqa
& &\bigcup_{m\geq 1}
\underline{\cal R}_{\rm WZ}(m,\varepsilon|p_{XY})
\nonumber\\
&\subseteq&{\cal R}_{\rm WZ}(\varepsilon|p_{XY})
\subseteq 
{\rm cl}\left(
\bigcup_{m \geq 1}
\underline{{\cal R}}_{\rm WZ}(m,\varepsilon|p_{XY})
\right).
\quad \label{eqn:aaaQw}
\eeqa
Note here that ${\cal R}_{\rm WZ}$$(\varepsilon$$|p_{XY})$ 
is a closed set. Then from (\ref{eqn:aaaQw}), we conclude that
\beqno
{\cal R}_{\rm WZ}(\varepsilon|W)
&= & 
{\rm cl}\left(
\bigcup_{m \geq 1}
\underline{{\cal R}}_{\rm WZ}(m,\varepsilon|p_{XY})
\right)
\\
&=&
{\rm cl}\left(
\bigcup_{m \geq 1}
\bigcap_{n \geq m}
{\cal R}_{\rm WZ}(n,\varepsilon|p_{XY})
\right),
\eeqno
completing the proof.
\hfill\IEEEQED
}
It is well known that $\tilde{\cal R}_{\rm WZ}(p_{XY})$ 
was determined by Wyner and Ziv \cite{wz}. 
To describe their result we introduce auxiliary random variables
$U$ and $Z$, respectively, taking values in finite sets 
${\cal U}$ and ${\cal Z}$. We assume that the joint 
distribution 
of $(U,X,Y,Z)$ is 
\beqno 
& &p_{U{X}{Y}\ByShun{Z}}(u,x,y,z)
\\
&=&p_{U}(u)p_{{X}|U}(x|u)p_{Y|X}(y|x)p_{Z|UY}(z|u,y).
\eeqno
The above condition is equivalent to 
$$
U \markov X \leftrightarrow Y,
X \markov (U,Y) \markov Z.
$$ 
Define the set of probability distribution $p=p_{UXYZ}$ by
\beqno
{\cal P}(p_{XY})
&\defeq &
\{p=p_{UXYZ}: \pa {\cal U} \pa \leq \pa {\cal X} \pa+1,
\\
& & U \markov  X \markov Y, X \markov (U,Y) \markov Z \},
\\
{\cal P}^{*}(p_{XY})
&\defeq &
\{p=p_{UXYZ}: \pa {\cal U} \pa \leq \pa {\cal X} \pa+1,
\\
& & U \markov X \markov Y, Z=\phi(U,Y)
\\
& &\mbox{ for some }
  \phi:{\cal U}\times {\cal Y}\to {\cal Z} \}.
\eeqno
By definitions it is obvious that 
${\cal P}^{*}(p_{XY})\subseteq {\cal P}(p_{XY})$.
Set
\beqno
{\cal R}(p)
&\defeq &
\ba[t]{l}
\{(R,\Dist): R, {\Dist} \geq 0\,,
\vSpa\\
\ba{rcl}
R & \geq & I_p(X;U|Y), {\Dist} \geq {\rm E}_p d(X,Z) \},
\ea
\ea
\\
{\cal R}(p_{XY})&\defeq& \bigcup_{p \in {\cal P}(p_{XY})}
{\cal R}(p),
\\
{\cal R}^{*}(p_{XY})&\defeq& \bigcup_{p \in {\cal P}^{*}(p_{XY})}
{\cal R}(p).
\eeqno
We can show that the above functions and sets satisfy 
the following property. 
\begin{pr}\label{pr:pro0}  
$\quad$
\begin{itemize}
\item[a)] 
The region ${\cal R}(p_{XY})$ is a closed convex set of 
$\mathbb{R}_{+}^2$.
\item[b)] For any $p_{XY}$, we have
$$
{\cal R}(p_{XY})={\cal R}^{*}(p_{XY}).
$$ 
\end{itemize}
\end{pr}

Proof of Property \ref{pr:pro0} is {\GorF} in Appendix \ref{sub:ApdaAAB}.
In Property \ref{pr:pro0} part b), 
${\cal R}(p_{XY})$ is regarded as another expression 
of ${\cal R}^{*}(p_{XY})$. This expression is useful 
for deriving our main result. 
%
%
\newcommand{\ApdaAAA}{
\subsection{Cardinality Bound on Auxiliary Random Variables}
\label{sub:ApdaAAA}


Define 
\beqno
{\cal Q}(p_{XY})
&\defeq& 
\{q_{UXYZ}: \pa {\cal U} \pa 
\leq \pa{\cal X}\pa\pa{\cal Y}\pa\pa{\cal Z}\pa,
q_{XY}=p_{XY},\\
&  & U \markov X \markov Y, X \markov (U,Y) \markov Z \}.
\eeqno
We first prove the following lemma.
\begin{lm}
\label{lm:CardLm}
\beqno
& &\underline{R}^{(\mu)}(p_{XY})
\defeq \min_{\scs q \in {\cal Q}(p_{XY})}
\left\{\bar{\mu}I_q(X;U|Y)+\mu {\rm E}_q d(X,Z)\right\}
\\
&=&R^{(\mu)}(p_{XY})
\defeq\min_{\scs q \in {\cal P}_{\rm sh}(p_{XY})}
\left\{\bar{\mu}I_q(X;U|Y)+\mu {\rm E}_q d(X,Z)\right\}.
\eeqno
\end{lm}

To prove Lemma \ref{lm:CardLm}, we set 
\beqno
& &{\cal Q}_{1}(p_{XY})
\defeq 
\{
\ba[t]{l}
q_1=q_{UXY}: |{\cal U}|\leq |{\cal X}||{\cal Y}||{\cal Z}|, 
\vspace*{1mm}\\ 
q_{XY}=p_{XY}, U \markov X \markov Y \}, 
\ea
\\
& &{\cal P}_{1}(p_{XY})
\defeq 
\{\ba[t]{l}
q_1=q_{UXY}:|{\cal U}|\leq |{\cal X}|, 
\vspace*{1mm}\\ 
q_{XY}=p_{XY}, U \markov X \markov Y \}, 
\ea
\\
& &{\cal Q}_{2}(q_{UXY})
 \defeq 
\{q_2=q_{Z|UXY}: 
\\
& & \qquad q_{UXYZ}=(q_{UXY},q_2),
    X \markov (U,Y)\markov Z\}.
\eeqno
By definition it is obvious that 
\beqa
{\cal Q}(p_{XY})&=&\{q=(q_1,q_2):
q_1\in {\cal Q}_1(p_{XY}),
\nonumber\\ 
& &q_2\in {\cal Q}_2(q_1)\},
\label{eqn:aax}\\
{\cal P}_{\rm sh}(p_{XY})&=&\{q=(q_1,q_2):
q_1\in {\cal P}_1(p_{XY}), 
\nonumber\\
& & q_2\in {\cal Q}_2(q_1)\}.
\label{eqn:aaxb}
\eeqa

{\it Proof of Lemma \ref{lm:CardLm}:} 
We first observe that by (\ref{eqn:aax}), 
we have
\beqno
& &\underline{R}^{(\mu)}(p_{XY})
\\
&=& 
\min_{q_1 \in {\cal Q}_1(p_{XY})}
\min_{q_2\in {\cal Q}_2(q_1)}
\left\{
\bar{\mu} I_{q_1}(X;U|Y)\right.
\\
& &\left.+\mu {\rm E}_{(q_1,q_2)}d(X,Z)
\right\}
\\
\\
&=& 
\min_{ q_1 \in {\cal Q}_1(p_{XY})}
\biggl\{
\bar{\mu} I_{q_1}(X;U|Y)
\\
& &\qquad \qquad \quad 
\left. +\mu 
\min_{ q_2\in {\cal Q}_2(q_1)}
{\rm E}_{(q_1,q_2)}d(X,Z)
\right\}
\\
&=& 
\min_{ q_1 \in {\cal Q}_1(p_{XY})}
\left\{
\bar{\mu} I_{q_1}(X;U|Y)+\mu 
{\rm E}_{(q_1,q^*_2(q_1))}d(X,Z)
\right\},
\eeqno
where 
$$
q_2^{*}=q^*_2(q_1)=q^*_{Z|UY}=\{q^*_{Z|UY}(z|u,y)\}_{
(u,y,z)\in {\cal U}\times {\cal Y}\times {\cal Z}}
$$
is a conditional probability distribution 
that attains the following optimization problem:
$$
\min_{q_2\in {\cal Q}_2(q_1)}
{\rm E}_{(q_1,q_2)}d(X,Z).
$$
We bound the cardinality $|{\cal U}|$ of $U$ 
to show that the bound $|{\cal U}|\leq |{\cal X}|$ is sufficient 
to describe $\underline{R}^{(\mu)}(p_{XY})$. Observe that 
\beqa
& &p_{X}(x)
=\sum_{u\in {\cal U}}q_U(u)
q_{X|U}(x|u),
\label{eqn:asdfq}
\\
& &\bar{\mu}I_{q_1}(X;U|Y)+ \mu {\rm E}_{(q_1,q^*_2)}(X,Z)
\nonumber\\
& &=\sum_{u\in {\cal U}}p_{U}(u)
\pi(q_{X|U}(\cdot|u)),
\label{eqn:aqqqaq}
\eeqa
where 
\beqno
&& \pi(q_{{X}|U}(\cdot|u))
\\
&\defeq& \sum_{(x,y,z)\in{\cal X} \times{\cal Y} \times{\cal Z}}
q_{X|U}(x|u)
p_{Y|X}(y|x)q^*_{Z|UY}(z|u,y)
\\
& & \times \log \left\{ \frac{q^{\bar{\mu}}_{X|U}(x|u)}
{p^{\bar{\mu}}_X(x)}
\frac{p^{\bar{\mu}}_{Y}(y){\rm e}^{-\mu d(x,z)}}
{\ds \left[\sum_{\tilde{x} \in {\cal X} }
p_{Y|X}(y|\tilde{x}) q_{X|U}(\tilde{x}|u)
\right]^{\bar{\mu}}}
\right\}.
\eeqno
For each  $u\in {\cal U}$, $\pi(q_{X|U}(\cdot|u))$ is a 
continuous function of $q_{X|U}(\cdot|u)$.
Then by the support lemma,
$$
|{\cal U}| \leq |{\cal X}|-1 +1= |{\cal X}| 
$$
is sufficient to express $|{\cal X}|-1$ values 
of (\ref{eqn:asdfq}) and 
one value of (\ref{eqn:aqqqaq}). 
\hfill \IEEEQED
} 
%
%
\newcommand{\ApdaAAB}{
\subsection{Proof of Property \protect{\ref{pr:pro0}}}
\label{sub:ApdaAAB}


In this appendix we prove Property \ref{pr:pro0}. Property \ref{pr:pro0} 
part a) is a well known property. Proof of this property is omitted here. 
We only prove Property \ref{pr:pro0} part b).

{\it Proof of Property \ref{pr:pro0} part b):} \ Since 
${\cal P}^*(p_{XY})$ $\subseteq $ ${\cal P}(p_{XY})$,
it is obvious that
${\cal R}^*(p_{XY})$ $\subseteq $ ${\cal R}(p_{XY})$. 
Hence it sufficies to prove that 
$
{\cal R}(p_{XY})$ $\subseteq $ ${\cal R}^*(p_{XY}). 
$
We assume that $(R,\Dist)\in {\cal R}(p_{XY})$. Then there exists 
$p \in {\cal P}(p_{XY})$ such that
\beq
R \geq I_p(U;X|Y)\mbox{ and }
{\Dist} \geq {\rm E}_p d(X,Z).
\label{eqn:sdR}
\eeq
On the second inequality in (\ref{eqn:sdR}), we have the following:   
\beqa
& &{\Dist}\geq {\rm E}_p d(X,Z)
\nonumber\\
&=&\sum_{\scs (u,y){\in {\cal U} \times {\cal Y}}}p_{UY}(u,y)
\left[
\sum_{z \in {\cal Z}} p_{Z|UY}(z|u,y)\right.
\nonumber\\
& &
\times 
\left.
\left(\sum_{x \in {\cal X}}d(x,z)p_{X|UY}(x|u,y)\right)
\right]
\nonumber\\
&\geq&
\sum_{\scs (u,y){\in {\cal U} \times {\cal Y}}}p_{UY}(u,y)
\nonumber\\
& &
\times
\left[
\min_{z \in {\cal Z}} 
\left(\sum_{x \in {\cal X}}d(x,z)p_{X|UY}(x|u,y)\right)
\right]
\nonumber\\
&=&
\sum_{\scs (u,y){\in {\cal U} \times {\cal Y}}}p_{UY}(u,y)
\left(\sum_{x \in {\cal X}}d(x,z^*)p_{X|UY}(x|u,y)\right),
\qquad
\label{eqn:zdF0001}
\eeqa
where $z^*=z^*(u,y)$ is one of the minimizers of the function 
$$
\sum_{x \in {\cal X}}d(x,z)p_{X|UY}(x|u,y).
$$
Define $\phi: {\cal U}\times {\cal Y}\to {\cal Z}$ by
$\phi(u,y)=z^*$. We further define  $q=q_{UXYZ}$ by 
$q_{UXY}=p_{UXY},q_{Z}=q_{\phi(U,Y)}.$ 
It is obvious that 
\beq 
q \in {\cal P}^*(p_{XY})\mbox{ and } 
R \geq I_p(X;U|Y) =I_q(X;U|Y). 
\label{eqn:Zs002}
\eeq 
Furthermore from (\ref{eqn:zdF0001}), we have
\beq
{\Dist}\geq {\rm E}_p d(X,Z) \geq {\rm E}_q d(X,\phi(U,Y)).
\label{eqn:Zs003}
\eeq
From (\ref{eqn:Zs002}) and (\ref{eqn:Zs003}), we have 
$(R,\Dist)\in {\cal R}^*(p_{XY})$. Thus
${\cal R}(p_{XY})$ $\subseteq $ 
${\cal R}^*(p_{XY})$ is proved.
\hfill\IEEEQED
}
The rate region ${\cal R}_{\rm WZ}(p_{XY})$ was determined by 
Wyner and Ziv \cite{wz}. Their result is the following.
\begin{Th}[Wyner and Ziv \cite{wz}]
\label{th:akw}
\begin{eqnarray*}
& &\tilde{\cal R}_{\rm WZ}(p_{XY})
={\cal R}^{*}(p_{XY})={\cal R}(p_{XY}).
\end{eqnarray*}
\end{Th}

On ${\cal R}_{\rm WZ}(p_{XY})$, Csisz\'ar and K\"orner 
\cite{ckBook81} obtained the following result. 
\begin{Th}[Csisz\'ar and K\"orner \cite{ckBook81}]
\label{th:ck}
\begin{eqnarray*}
& &{\cal R}_{\rm WZ}(p_{XY})=\tilde{\cal R}_{\rm WZ}(p_{XY})
={\cal R}^*(p_{XY})={\cal R}(p_{XY}).
\end{eqnarray*}
\end{Th}

We are interested in an asymptotic behavior of the error probability 
of decoding to tend to one as $n \to \infty$ for 
$(R,\Dist) \notin {\cal R}_{\rm WZ}(p_{XY})$. 
To examine the rate of convergence, we define the 
following quantity. 
Set
\beqno
& & {\rm P}_{\rm c}^{(n)}(\varphi^{(n)},\psi^{(n)};{\Dist})
\defeq 
1-{\rm P}_{\rm e}^{(n)}(\varphi^{(n)},\psi^{(n)};{\Dist}),
\\
& & 
G^{(n)}(R,\Dist|p_{XY})
\\
&&\defeq
\min_{\scs 
(\varphi^{(n)},\psi^{(n)}):
    \atop{\scs 
    (1/n)\log \|\varphi^{(n)}\|\leq R 
    }
}
\hspace*{-4mm}
\left(-\frac{1}{n}\right)
\log {\rm P}_{\rm c}^{(n)}(\varphi^{(n)},\psi^{(n)};{\Dist}).
\eeqno
By time sharing we have that 
\beqa
& &G^{(n+m)}\left(\left.
\frac{n R+m R^{\prime}}{n+m},
\frac{n {\Dist} + m{\Dist}^{\prime}}{n+m}\right|p_{XY}\right) 
\nonumber
\\
&\leq& \frac{nG^{(n)}(R,\Dist|p_{XY}) 
+mG^{(m)}(R^{\prime},{\Dist}^{\prime}|p_{XY})}{n+m}. 
\label{eqn:aaZ} 
\eeqa
Choosing $R=R^\prime$ and $\Dist=\Dist^\prime$ in (\ref{eqn:aaZ}), we 
obtain the following subadditivity property
on $\{G^{(n)}(R,\Dist|p_{XY})$ $\}_{n\geq 1}$: 
\beqno
& &G^{(n+m)}(R,\Dist|p_{XY}) 
\\
&\leq& \frac{nG^{(n)}(R,\Dist|p_{XY}) 
+mG^{(m)}(R,\Dist|p_{XY})}{n+m},
\eeqno
which together with Fekete's lemma yields 
that $G^{(n)}(R,$ $\Dist|p_{XY})$ exists 
and satisfies the following: 
\beqno
\lim_{n\to\infty}G^{(n)}(R,\Dist|p_{XY}) 
=\inf_{n\geq 1}G^{(n)}(R,\Dist|p_{XY}).
\eeqno
Set 
\beqno
& & G(R,\Dist|p_{XY})\defeq
\lim_{n\to\infty}G^{(n)}(R,\Dist|p_{XY}),
\\
& &{\cal G}(p_{XY})\defeq \{
(R,\Delta,G): G \geq G(R,\Dist|p_{XY})\}. 
\eeqno
The exponent function $G(R,\Dist|p_{XY})$ is a convex function 
of $(R,\Dist)$. In fact, from (\ref{eqn:aaZ}), we have that  
for any $\alpha \in [0,1]$
\beqno
& &G(\alpha R+\bar{\alpha}R^{\prime},
     \alpha {\Dist}+\bar{\alpha}{\Dist}^{\prime}|p_{XY})
\\
&\leq &
\alpha G(R,\Dist|p_{XY})
+\bar{\alpha} G( R^{\prime},{\Dist}^{\prime}|p_{XY}).
\eeqno
The region ${\cal G}(p_{XY})$ is also 
a closed convex set. Our main aim is to 
find an explicit characterization of ${\cal G}(p_{XY})$. In 
this paper we derive an explicit outer bound of ${\cal G}$ 
$(p_{XY})$ whose section by the plane $G=0$ coincides 
with ${\cal R}_{\rm WZ}(p_{XY})$.

\section{Main Result}
%
%

In this section we state our main result. 
We first  explain that the rate distortion region 
${\cal R}(p_{XY})$ can be expressed 
with two families of supporting hyperplanes.
To describe this result we define two sets of probability 
distributions on ${\cal U}$ $\times{\cal X}$ 
$\times{\cal Y}$ by
\beqno
{\cal P}_{\rm sh}(p_{XY})
&\defeq &
\{p_{UXYZ}: \pa {\cal U} \pa \leq \pa {\cal X} \pa,
U \markov  X \markov Y,\\
& & X \markov (U,Y) \markov Z \}.
\\
{\cal Q}
&\defeq& 
\{q=q_{UXYZ}:\pa {\cal U} \pa \leq \pa {\cal X} \pa
\pa {\cal Y} \pa \pa {\cal Z} \pa \}.
\eeqno
We set
\beqno
& & R^{(\mu)}(p_{XY})
\\
& \defeq& 
\max_{p \in {\cal P}_{\rm sh}(p_{XY})}
\left\{\bar{\mu} I_p(X;U|Y)+ \mu {\rm E}_p d(X;Z)\right\},
\eeqno
where $\bar{\mu}=1-\mu$. Furthermore set
\beqno
& & \tilde{R}^{(\alpha, \mu)}(p_{XY})
\\
&\defeq& 
\min_{
{\scs q\in {{\cal Q}}
}}
\ba[t]{l}
\left\{
 \bar{\alpha}[ D(q_X||p_X)
+ D(q_{Y|XU}||p_{Y|X}|q_{XU})
\right.
\vspace{1mm} \\
\left. \hspace*{2mm}+I_q(X;Z|UY)]
\right.
\vspace{1mm} \\
\left. \hspace*{2mm} 
+ 4\alpha[\bar{\mu}\ByShun{I_q(X;U|Y)} + \mu {\rm E}_q d(X,Z)]\right\},
\ea
\\
&&{\cal R}_{\rm sh}(p_{XY})
\\
& \defeq &\bigcap_{\mu \in [0,1]} \{(R,\Dist): 
\bar{\mu} R+\mu {\Dist} \geq R^{(\mu)}(p_{XY})\},
\\
& & \tilde{\cal R}_{\rm sh}^{(\alpha)}(p_{XY})
\\
&\defeq& 
\bigcap_{\scs \mu\in [0,1]
        }
\left\{(R,\Dist): \bar{\mu} R + \mu {\Dist} \geq 
\frac{1}{4\alpha}\tilde{R}^{(\alpha, \mu)}(p_{XY}) \right\},
\\
& & \tilde{\cal R}_{\rm sh}(p_{XY})
\defeq \bigcap_{\alpha \in (0,1]} \tilde{\cal R}_{\rm sh}^{(\alpha)}(p_{XY})
\\
&=& 
\bigcap_{\scs \mu\in [0,1], 
        \atop{\scs \alpha \in (0,1]}
        }
\left\{(R,\Dist): \bar{\mu} R+ \mu {\Dist} \geq 
\frac{1}{4\alpha}\tilde{R}^{(\alpha, \mu)}(p_{XY}) \right\}.
\eeqno
For ${\cal R}\subseteq \mathbb{R}_{+}^2$, we set
$$
{\cal R}-\kappa(1,1)
\defeq \{(a-\kappa,b-\kappa)\in \mathbb{R}_{+}^2: (a,b)\in{\cal R}\}.
$$
Then we have the following property.
\begin{pr}
\label{pr:pro0z} 
For any $p_{XY}$, we have
\beq
{\cal R}_{\rm sh}(p_{XY})={\cal R}(p_{XY}).
\label{eqn:PropEqB}
\eeq
For any $\alpha\in (0,\alpha_0]$, 
we have
\beqa
& &{R}^{(\mu)}(p_{XY})
-c_1 \sqrt{\frac{\alpha}{\bar{\alpha}}}
\log\left(c_2\frac{\bar{\alpha}}{\alpha}\right) 
\nonumber\\
&\leq & \frac{1}{4\alpha}\tilde{R}^{(\alpha,\mu)}(p_{XY})
\leq {R}^{(\mu)}(p_{XY}),
\label{eqn:PropEqA}
\eeqa
where
\beq
\left.
\ba{rcl}
\alpha_0&=& \alpha_0(d_{\max},|{\cal X}|)
\defeq [32\log(|{\cal X}|{\rm e}^{d_{\max}})+1]^{-1},
\vspace{1mm}\\
c_1&=& c_1(d_{\max},|{\cal X}|)
\defeq 
4\sqrt{2\log(|{\cal X}|{\rm e}^{d_{\max}})},
\vspace{1mm}\\
c_2&=&\ds c_2(d_{\max}, |{\cal X}|,|{\cal Y}|,|{\cal Z}|)
\defeq 
\frac{{\rm e}^{ \frac{1}{2} d_{\max}} 
       |{\cal Z}| |{\cal X}|^2 |{\cal Y}|^3} 
{8\log(|{\cal X}|{\rm e}^{d_{\max}})}.
\ea
\right\}
\label{eqn:AlpaConeCtwo}
\eeq
The two inequalities of (\ref{eqn:PropEqA}) implies that 
for each $\alpha\in (0,\alpha_0]$, 
\beqa
& &{\cal R}_{\rm sh}(p_{XY})-c_1 \sqrt{\frac{\alpha}{\bar{\alpha}}}
\log\left(c_2\frac{\bar{\alpha}}{\alpha}\right)(1,1) 
\nonumber\\
&\subseteq& 
\tilde{\cal R}_{\rm sh}^{(\alpha)}(p_{XY})
\subseteq {\cal R}_{\rm sh}(p_{XY}).
\nonumber 
\eeqa
Hence we have
\beqa
\tilde{\cal R}_{\rm sh}(p_{XY})={\cal R}_{\rm sh}(p_{XY}).
\eeqa
\end{pr}

Proof of Property \ref{pr:pro0z} is {\GorF} in Appendix \ref{sub:ApdaAABz}.
\newcommand{\ApdaAABz}{
\subsection{Proof of Property \protect{\ref{pr:pro0z}}}
\label{sub:ApdaAABz}


In this appendix we prove Property \ref{pr:pro0z}. 
From Property \ref{pr:pro0} part a), we have 
the following lemma. 
\begin{lm}\label{lm:asgsq} 
Suppose that  
$(\hat{R},\hat{\Dist})$ does not belong to ${\cal R}($ $p_{XY})$. 
Then there exist $\epsilon, \mu^*>0$ such that 
for any $(R,\Dist)\in {\cal R}(p_{XY})$
we have  
\beqno
\bar{\mu}(R-\hat{R})+\mu^*({\Dist}-\hat{\Dist})-\epsilon \geq 0.
\eeqno
\end{lm} 

Proof of this lemma is omitted here. Lemma \ref{lm:asgsq} is 
equivalent to the fact that if the region ${\cal R}(p_{XY})$ 
is a convex set, then for any point $(\hat{R},\hat{\Dist})$ 
outside the region ${\cal R}(p_{XY})$, there exits 
a line which separates the point $(\hat{R},\hat{\Dist})$ 
from the region ${\cal R}(p_{XY})$. 
Lemma \ref{lm:asgsq} will be used to prove 
(\ref{eqn:PropEqB}) in Property \ref{pr:pro0z}.

{\it Proof of (\ref{eqn:PropEqB}) in Property \ref{pr:pro0z}:} \ 
We first recall the following definitions 
of ${\cal P}(p_{XY})$ and ${\cal P}_{\rm sh}(p_{XY})$: 
\beqno
{\cal P}(p_{XY})
&\defeq& 
\{p_{UXYZ}: \pa {\cal U} \pa \leq \pa {\cal X}\pa+1, 
U \markov  X \markov Y,
\\
& & 
X \markov  (U,Y) \markov Z \},
\\
{\cal P}_{\rm sh}(p_{XY})
&\defeq& 
\{
p_{UXYZ}: \pa {\cal U} \pa \leq \pa{\cal X}\pa,
U \markov  X\markov Y, 
\\
& & X \markov  (U,Y) \markov Z \}.
\eeqno
We prove 
${\cal R}_{\rm sh}(p_{XY})$ $\subseteq {\cal R}(p_{XY})$. 
We assume that $(\hat{R},\hat{\Dist})\notin {\cal R}(p_{XY})$. 
Then by Lemma \ref{lm:asgsq}, there exist 
$\epsilon>0$ and  $\mu^*>0$ 
such that for any $(R,\Dist)\in {\cal R}(p_{XY})$, we have
\beqno
\bar{\mu}^* \hat{R}+\mu^* \hat{\Dist}\leq \bar{\mu}^* R
+ \mu^* \Dist-\epsilon.
\eeqno
Hence we have
\beqa
& &\bar{\mu}^* \hat{R}+\mu^* \hat{\Dist}
\leq 
   \min_{(R,\Dist)\in {\cal R}(p_{XY})} 
   \left\{ \bar{\mu}^{\ast} R+ \mu^* \Dist \right\} 
   -\epsilon
\nonumber\\ 
&\MEq{a}&\min_{p\in {\cal P}(p_{XY})}
\left\{ \bar{\mu}^* I_p(U;X|Y)+ \mu^* {\rm E}_pd(X,Z) \right\}-\epsilon
\nonumber\\ 
&\leq&\min_{p\in {\cal P}_{\rm sh}(p_{XY})}
\left\{ \bar{\mu}^* I_p(U;X|Y)+ \mu^* {\rm E}_pd(X,Z) \right\}-\epsilon
\nonumber\\ 
&=&R^{(\mu^*)}(p_{XY})-\epsilon.
\label{eqn:sddsd}
\eeqa
Step (a) follows from the definition of ${\cal R}(p_{XY})$. 
The inequality (\ref{eqn:sddsd}) implies that 
$(\hat{R},\hat{\Dist})$ $\notin {\cal R}_{\rm sh}(p_{XY})$.
Thus ${\cal R}_{\rm sh}(p_{XY})$ $\subseteq {\cal R}(p_{XY})$ 
is concluded. 
We next prove 
${\cal R}($ $p_{XY})$ $\subseteq $ 
${\cal R}_{\rm sh}$$(p_{XY})$. 
We assume that $(R,$ ${\Dist}) \in$ ${\cal R}(p_{XY})$.
Then there exists $q\in$ ${\cal P}$ $(p_{XY})$ such that
\beq
R\geq I_q(X;U|Y), \Dist \geq {\rm E}_q d(X,Z).
\label{eqn:assz0}
\eeq
Then, for each $\mu>0$ and for $(R,{\Dist})$ $\in {\cal R}(p_{XY})$, 
we have the following chain of inequalities:
\beqno
& &\bar{\mu} R+\mu{\Dist} \MGeq{a} \bar{\mu}I_q(X;U|Y) + \mu {\rm E}_q d(X,Z)
\\
&\geq & 
\min_{q\in {\cal P}(p_{XY})}
\left[\bar{\mu} I_q(X;U|Y) + \mu {\rm E}_q d(X,Z) \right]
\\
&=&{R}^{(\alpha,\mu)}(p_{XY}).
\eeqno
Step (a) follows from (\ref{eqn:assz0}).
Hence we have 
${\cal R}(p_{XY})$ $\subseteq $ 
${\cal R}_{\rm sh}(p_{XY})$.
\hfill\IEEEQED

We next prove the two inequalities of 
(\ref{eqn:PropEqA}) in  Property \ref{pr:pro0z}.

{\it Proof of (\ref{eqn:PropEqA}) in Property \ref{pr:pro0z}:}
We first prove the second inequality of 
(\ref{eqn:PropEqA}) in  Property \ref{pr:pro0z}.
We have the following chain of inequalities.
\beqno
& & 
{R}^{(\alpha,\mu)}(p_{XY}).
\\
&\geq & 
\min_{q\in {\cal P}(p_{XY})}
\left[\bar{\mu} I_q(X;U|Y) + \mu {\rm E}_q d(X,Z) \right]
\nonumber\\
&\MEq{a}&\min_{\scs 
  \atop{\scs 
  q\in {\cal P}(p_{XY})
  }
}
\ba[t]{l}
\left\{ \ds 
 \frac{\bar{\alpha}}{4\alpha}
[ D(q_X||p_X) + D(q_{Y|XU}||p_{Y|X}|q_{XU})\right.
\vspace{1mm}\\
\hspace*{2mm}+\gamma I_q(X;Z|UY)]
\vspace{1mm}\\
\hspace*{2mm} + \bar{\mu} I_q(X;U|Y) + \mu {\rm E}_q d(X,Z) \Bigr\}
\ea
\nonumber\\
&\geq&\min_{\scs 
  \atop{\scs 
  q\in {\cal Q}
  }
}
\ba[t]{l}
\left\{\ds 
 \frac{ \bar{\alpha}}{4\alpha}[D(q_X||p_X) + D(q_{Y|XU}||p_{Y|X}|q_{XU})\right.
\vspace{1mm}\\
\hspace*{2mm}+\gamma I_q(X;Z|UY)]
\vspace{1mm}\\
\hspace*{2mm}  + \bar{\mu} I_q(X;U|Y) + \mu {\rm E}_q d(X,Z) \Bigr\}
\ea
\\
&=&\frac{1}{4\alpha} \tilde{R}^{(\alpha,\mu)}(p_{XY}).
\eeqno
Step (a) follows from that 
when $q \in {\cal P}(p_{XY})$, we have
$$
D(q_{X}|| p_{X})=D(q_{Y|XU}||p_{Y|X}|q_{XU})
=I_q(X;Z|UY)=0,
$$
We next prove the first inequality of (\ref{eqn:PropEqA}) in 
Property \ref{pr:pro0z}. Let 
$
q^*_{\alpha,\mu}=q^*_{UXYZ,\alpha,\mu}
\in {\cal Q}
$
be a probability distribution which attains the minimum 
in the definition of $\tilde{R}^{(\alpha,\mu)}(p_{XY})$.
Let
$\hat{q}_{\alpha,\mu}$
$=\hat{q}_{UXYZ,}$${}_{\alpha,\mu}$ be 
a probability distribution with the form
\beqno  
& &\hat{q}_{UXYZ,\alpha,\mu}(u,x,y,z)
\\
&=&q_{U|X,\alpha,\mu}^*(u|x)
   p_{X}(x)p_{Y|X}(y|x)q_{Z|UY,\alpha,\mu}^*(z|u,y).
\eeqno
Define
\beqno
{\cal Q}(p_{XY})
&\defeq& 
\{q_{UXYZ}: \pa {\cal U} \pa 
\leq \pa{\cal X}\pa\pa{\cal Y}\pa\pa{\cal Z}\pa,
q_{XY}=p_{XY},\\
&  & U \markov X \markov Y, X \markov (U,Y) \markov Z \}.
\eeqno
By definition, we have
$\hat{q}_{\alpha,\mu}
\in {\cal Q}(p_{XY})$. 
Then we have the following chain of inequalities.
\beqa
&&\bar{\alpha}D(q^*_{\alpha,\mu}||
\hat{q}_{\alpha,\mu})
\nonumber\\
&=&
\bar{\alpha}[D(q^*_{X,\alpha,\mu}||p_{X})
+ D(q^*_{Y|XU,\alpha,\mu}||
       p_{Y|X}|q^*_{XU,\alpha,\mu})
\nonumber\\
& & + I_{q^*_{\alpha,\mu}}(X;Z|UY)]
\nonumber\\
&\leq &\bar{\alpha}[D(q_{X,\alpha,\mu}^*||p_X)
+ D(q_{Y|XU,\alpha,\mu}^*||p_{Y|X}|q^*_{XU,\alpha,\mu})
\nonumber\\
& &    +\bar{\mu}I_{q^*_{\alpha,\mu}}(X;Z|UY)]
\nonumber\\
& &   + 4\alpha
       [ \bar{\mu} I_{q^*_{\alpha, \mu}}(X;U|Y)
      + \mu {\rm E}_{q^*_{\alpha,\mu}}d(X;Z)]
\nonumber\\
&=& 
\tilde{R}^{(\alpha,\mu)}(p_{XY})
\nonumber\\
&\leq &
\bar{\alpha}[D(\hat{q}_{X,\alpha,\mu}||p_X)
    + D(\hat{q}_{Y|XU,\alpha,\mu}||p_{Y|X}|\hat{q}_{XU,\alpha,\mu})
\nonumber\\
& &    + I_{\hat{q}_{\alpha,\mu}}(X;Z|UY)]
\nonumber\\
& &   +  4\alpha[\bar{\mu} I_{\hat{q}_{\alpha, \mu}}(X;U|Y)
      + \mu {\rm E}_{\hat{q}_{\alpha,\mu}}d(X;Z)]
\nonumber\\
&=&  4\alpha[\bar{\mu} I_{\hat{q}_{\alpha, \mu}}(X;U|Y)
    + \mu {\rm E}_{\hat{q}_{\alpha,\mu}}d(X;Z)]
\nonumber\\
&\leq & 4\alpha\log(|{\cal X}|{\rm e}^{d_{\max}}).
\label{eqn:zxaw}
\eeqa
For simplicity of notation we set
$
\xi\defeq \log(|{\cal X}|{\rm e}^{d_{\max}}).
$
From (\ref{eqn:zxaw})
we have 
\beq
D(q^*_{\alpha,\mu}||\hat{q}_{\alpha,\mu}) \leq 
4\xi \frac{\alpha}{\bar{\alpha}}.
\label{eqn:zxawX}
\eeq
By the Pinsker's inequality we have 
\beq
\frac{1}{2}(||q^*_{\alpha,\mu}-\hat{q}_{\alpha,\mu}||_1)^2
\leq  D(q^*_{\alpha,\mu}||\hat{q}_{\alpha,\mu}) 
\label{eqn:zxawXz}
\eeq
From (\ref{eqn:zxawX}) and (\ref{eqn:zxawXz}), 
we obtain
\beq
||q^*_{\alpha,\mu}-\hat{q}_{\alpha,\mu}||_1
\leq \sqrt{8\xi \frac{\alpha}{\bar{\alpha}}}\leq \frac{1}{2}
\label{eqn:zxawXzZ}
\eeq
for any $\alpha \in (0, \alpha_0]$ with 
\beq
\alpha_0=(32\xi+1)^{-1}
=[32\log (|{\cal X}|{\rm e}^{d_{\max}})+1]^{-1}.
\eeq
The bound (\ref{eqn:zxawXzZ}) implies that for any 
$A \subseteq \{U,X,Y,Z\}$, we have 
\beq
||q^*_{A,\alpha,\mu}-\hat{q}_{A,\alpha,\mu}||_1
\leq \sqrt{8\xi \frac{\alpha}{\bar{\alpha}}}\leq \frac{1}{2}.
\label{eqn:zxawXzZzz}
\eeq
Then for any $\alpha \in (0, \alpha_0]$, 
we have the following chain of inequalities:
\beqa
&&|I_{q^*_{\alpha,\mu}}(U;X|Y)-I_{\hat{q}_{\alpha,\mu}}(U;X|Y)|
\nonumber\\
&\leq & 
 |H_{q^*_{\alpha,\mu}}(XY)-H_{\hat{q}_{\alpha,\mu}}(XY)|
\nonumber\\
& &
+|H_{q^*_{\alpha,\mu}}(Y)-H_{\hat{q}_{\alpha,\mu}}(Y)|
\nonumber\\
& &
+|H_{q^*_{\alpha,\mu}}(UXY) - H_{\hat{q}_{\alpha,\mu}}(UXY)|
\nonumber\\
& &
+|H_{q^*_{\alpha,\mu}}(UY)- H_{\hat{q}_{\alpha,\mu}}(UY)|
\nonumber\\
&\MLeq{a} &
-\sqrt{8\xi \frac{\alpha}{\bar{\alpha}}}
\left[
\log
\left(
\sqrt{8\xi \frac{\alpha}{\bar{\alpha}}}
\cdot \frac{1}{ |{\cal X}| |{\cal Y}| }
\right)\right.
\nonumber\\
& &
+\log
\left(
\sqrt{8\xi \frac{\alpha}{\bar{\alpha}}}
\cdot \frac{1}{|{\cal Y}|}
\right)
+\log
\left(
\sqrt{8\xi \frac{\alpha}{\bar{\alpha}}}
\cdot \frac{1}{ |{\cal U}||{\cal X}| |{\cal Y}|}
\right)
\nonumber\\
& &\left. 
+\log
\left(
\sqrt{8\xi \frac{\alpha}{\bar{\alpha}}}
\cdot \frac{1}
{|{\cal U}| |{\cal Y}|}
\right)
\right]
\nonumber\\
&=&
2\sqrt{8\xi \frac{\alpha}{\bar{\alpha}}}
\log\left\{
\left(\frac{\bar{\alpha}}{8\xi{\alpha}}\right)
|{\cal U}||{\cal X}||{\cal Y}|^2
\right\}
\nonumber\\
&\MLeq{b}&
2\sqrt{8\xi \frac{\alpha}{\bar{\alpha}}}
\log\left\{
\left(\frac{\bar{\alpha}}{8\xi{\alpha}}\right)
|{\cal Z}||{\cal X}|^2|{\cal Y}|^3
\right\}.
\label{eqn:DatZz}
\eeqa
Step (a) follows from (\ref{eqn:zxawXzZzz}) and 
LEMMA 2.7 in Section 1.2 in Csisz\'ar and K\"orner \cite{ckBook81}. 
Step (b) follows from that when $q\in {\cal Q}$,
we have $|{\cal U}|\leq |{\cal X}|$$|{\cal Y}| |{\cal Z}|$.
On the other hand we have  
\beq
|{\rm E}_{q^*_{\alpha,\mu}}d(X,Z)
-{\rm E}_{\hat{q}_{\alpha,\mu}}d(X,Z)|
\leq d_{\max} 
\sqrt{8\xi \frac{\alpha}{\bar{\alpha}}}.
\label{eqn:DatZza}
\eeq
Combining (\ref{eqn:DatZz}) and (\ref{eqn:DatZza}), 
we have 
\beqa
& & \left|
  \bar{\mu}I_{q^*_{\alpha,\mu}}(U;X|Y)
+\mu {\rm E}_{q^*_{\alpha,\mu}}d(X,Z)\right.
\nonumber\\
& & \left. -\left(
\bar{\mu}I_{\hat{q}_{\alpha,\mu}}(U;X|Y)
+\mu {\rm E}_{\hat{q}_{\alpha,\mu}}d(X,Z)\right)
\right|
\nonumber\\
&\leq & 
|I_{q^*_{\alpha,\mu}}(U;X|Y)-I_{\hat{q}_{\alpha,\mu}}(U;X|Y)|
\nonumber\\
& & +|{\rm E}_{q^*_{\alpha,\mu}}d(X,Z)
-{\rm E}_{\hat{q}_{\alpha,\mu}}d(X,Z)|
\nonumber\\
&\leq &
2\sqrt{8\xi \frac{\alpha}{\bar{\alpha}}}
\log\left\{
\left(\frac{\bar{\alpha}}{8\xi{\alpha}}\right)
|{\cal Z}||{\cal X}|^2|{\cal Y}|^3
\right\}
+d_{\max} 
\sqrt{8\xi \frac{\alpha}{\bar{\alpha}}}
\nonumber\\
&=&c_1 \sqrt{\frac{\alpha}{\bar{\alpha}}}
\log\left(c_2\frac{\bar{\alpha}}{\alpha}\right). 
\eeqa
Then we have 
\beqa
&& \frac{1}{4\alpha}
\tilde{R}^{(\alpha,\mu)}(p_{XY})
\nonumber\\
&= &\frac{\bar{\alpha}}{4\alpha} 
[D(q_{X,\alpha,\mu}^*||p_X)
+ D(q_{Y|XU,\alpha,\mu}^*||p_{Y|X}|q^*_{XU,\alpha,\mu})
\nonumber\\
& &    + I_{q^*_{\alpha,\mu}}(X;Z|UY)]
\nonumber\\
& &   +  I_{q^*_{\alpha, \mu}}(X;U|Y)
      + \mu {\rm E}_{q^*_{\alpha,\mu}}d(X;Z)
\nonumber\\
&\geq &  I_{q^*_{\alpha, \mu}}(X;U|Y)
        + \mu {\rm E}_{q^*_{\alpha,\mu}}d(X;Z)
\nonumber\\
&\geq & I_{\hat{q}_{\alpha, \mu}}(X;U|Y)
        + \mu {\rm E}_{\hat{q}_{\alpha,\mu}}d(X;Z)
\nonumber\\
&  & -c_1 \sqrt{\frac{\alpha}{\bar{\alpha}}}
\log\left(c_2\frac{\bar{\alpha}}{\alpha}\right)
\nonumber\\
&\MGeq{a}& 
\min_{\scs q \in {\cal Q}(p_{XY})}
\left\{I_q(X;Y|U)+\mu {\rm E}_{q}d(X,Z)\right\}
\nonumber\\
& &-c_1 \sqrt{\frac{\alpha}{\bar{\alpha}}}
\log\left(c_2\frac{\bar{\alpha}}{\alpha}\right)
\nonumber\\
&\MEq{b}& 
\min_{\scs q \in {\cal P}_{\rm sh}(p_{XY})}
\left\{I_q(X;Y|U) +\mu {\rm E}_q d(X,Z) \right\} 
\nonumber\\
& &-c_1 \sqrt{\frac{\alpha}{\bar{\alpha}}}
\log\left(c_2\frac{\bar{\alpha}}{\alpha}\right)
\nonumber\\
&=&{R}^{(\mu)}(p_{XY}) 
-c_1 \sqrt{\frac{\alpha}{\bar{\alpha}}}
\log\left(c_2\frac{\bar{\alpha}}{\alpha}\right).
\label{eqn:aafgz}
\eeqa
Step (a) follows from that 
$\hat{q}_{\alpha,\mu}$ 
$\in {\cal Q}(p_{XY})$.
Step (b) follows from 
Lemma \ref{lm:CardLm} stated in Appendix \ref{sub:ApdaAAA}.
\hfill\IEEEQED
}
For $ (\alpha,\mu) \in [0,1]^2$, define
\beqno
& &\omega_{\ByShun{q||p}}^{(\alpha, \mu)}(x,y,z|u)
\\
&\defeq&
 \bar{\alpha}\left [\log \frac{q_{{X}}(x)}{p_{X}(x)}
  +\log \frac{q_{Y|XU}(y|x,u)}{p_{Y|X}(y|x)}\right.
\\
& &
\left.
+\log \frac{q_{X|UYZ}(x|u,y,z)}{q_{X|UY}(x|u,y)}\right]
\\
& &+4\alpha\left[\log \frac{q_{X|YU}(x|u,y)}{p_{X|Y}(x|y)}
 +\mu d(x,z)\right],
\\
& &
f^{(\alpha, \mu,\lambda)}_{{\pANDq}}(x,y,z|u)
\defeq
\exp\left\{
-\lambda
\omega^{(\alpha,\mu)}_{{\pANDq}}(x,y,z|u)
\right\},
\\
& &\Omega^{(\alpha,\mu,\lambda)}(q|p_{XY})
\\
&\defeq& 
-\log 
{\rm E}_{q}
\left[\exp\left\{-\lambda
\omega^{(\alpha,\mu)}_{{\pANDq}}(X,Y,Z|U)\right\}\right]
\\
&=& 
-\log 
\left[
\sum_{u,x,y,z}q(u,x,y,z)
f^{(\alpha, \mu, \lambda)}_{{\pANDq}}(x,y,z|u)
\right],
\\
& &
\Omega^{(\alpha,\mu, \lambda)}(p_{XY})
\defeq 
\min_{\scs \atop{\scs 
q \in {{\cal Q}}
}}
\newcommand{\zaP}{
%
%
\min_{
\scs (\tilde{U},\tilde{X},\tilde{Y})
\in {\cal U} \times{\cal X}\times{\cal Y}:
       \atop{\scs \tilde{U} \leftrightarrow 
          \tilde{X}\leftrightarrow \tilde{Y}, 
            |{\cal U}| \leq |{\cal X}|,
           \atop{\scs
            p_{\tilde{Y}\tilde{X}|\tilde{U}}=p_{YX|U}
           } 
       }
    }
}
\Omega^{(\alpha,\mu, \lambda)}(q|p_{XY}),
\\
& &F^{(\alpha,\mu, \lambda)}(\bar{\mu} R+ \mu{\Dist}|p_{XY})
\\
&\defeq&
\frac{\Omega^{(\alpha,\mu,\lambda)}
(p_{XY})-4\alpha\lambda( \bar{\mu}R+\mu{\Dist})}
{1+4(1-\alpha\mu)\lambda},
\\
& &F(R,\Dist|p_{XY})
\\
&\defeq& \sup_{(\alpha,\mu)\in (0,1]^2, \lambda>0} 
F^{(\alpha, \mu,\lambda)}(\bar{\mu} R+\mu{\Dist}|p_{XY}),
\\
& &\overline{\cal G}(p_{XY})
\defeq \left\{(R,\Dist,G): G\geq F(R,\Dist|p_{XY})
\right\}.
\eeqno
We can show that the above functions and sets satisfy 
the following property.
\begin{pr}\label{pr:pro1}  
$\quad$
\begin{itemize}

\item[a)] 
The cardinality bound 
$|{\cal U}|\leq|{\cal X}|$$|{\cal Y}|$$|{\cal Z}|$
appearing in  
the definition of 
$\Omega^{(\alpha, \mu,\lambda)}(p_{XY})$ 
is sufficient to describe this quantity.
\item[b)] 
Define a probability distribution 
$q^{(\lambda)}=
q_{UXYZ}^{(\lambda)}
$
by
\beqno 
& &q^{(\lambda)}(u,x,y,z)
\\
&\defeq&
\frac{
q(u,x,y,z)
\exp\left\{-\lambda
\omega^{(\alpha,\mu)}_{{\pANDq}}(x,y,z|u)\right\}
}
{
{\rm E}_{q}
\left[\exp\left\{-\lambda
\omega^{(\alpha, \mu)}_{{\pANDq}}(X,Y,Z|U)\right\}\right]
}.
\eeqno
Then we have 
\beqno
& & \frac{\rm d}{{\rm d}\lambda} \ByShun{\Omega^{(\alpha, \mu, \lambda)}}(q|p_{XY})
={\rm E}_{q^{(\lambda)}}
\left[
\omega^{(\alpha, \mu)}_{{\pANDq}}(X,Y,Z|U)
\right],
\\
& & \frac{\rm d^2}{{\rm d}\lambda^2} \ByShun{\Omega^{(\alpha, \mu, \lambda)}}(q|p_{XY})
\\
& &\qquad =-{\rm Var}_{q^{(\lambda)}}
\left[
\omega^{(\alpha, \mu)}_{{\pANDq}}(X,Y,Z|U)\right].
\eeqno
The second equality implies that 
$\Omega^{(\alpha, \mu, \lambda)}_q(q|p_{XY})$ 
is a concave function of $\lambda>0$. 
\item[c)]
Define 
\beqno
& &\rho =\rho(p_{XY})
\\
& &\defeq
\max_{q \in {\cal Q}}
\max_{\scs (\alpha,\mu)\in [0,1]^2} 
{\rm Var}_{q}
\left[\omega^{(\alpha, \mu)}_{{\pANDq}}(X,Y,Z|U)\right].
\eeqno
Since 
$$
0\leq \left[\omega^{(\alpha, \mu)}_{\pANDq}(x,y,z|u)\right]^2< \infty 
$$
for $(u,x,y,z)
\in {\cal U} \times{\cal X} \times{\cal Y} \times{\cal Z}$,
we have $\rho(p_{XY})<\infty $.
Then for any $\lambda \in (0,1]$, we have
\beqno
& &\Omega^{(\alpha, \mu, \lambda)}(q|p_{XY}) 
\\
&\geq& \lambda {\rm E}_{q}
\left[\omega^{(\alpha, \mu)}_{{\pANDq}}(X,Y,Z|U)\right]
-\frac{1}{2}\rho(p_{XY})\lambda^2
\\
&=& 
\ba[t]{l}
 \bar{\alpha}[ D(q_X||p_X)
+ D(q_{Y|XU}||p_{Y|X}|q_{XU})
\vspace{1mm}\\
\hspace*{2mm}+I_q(X;Z|UY)]
+ 4\alpha[\bar{\mu}\ByShun{I_q(X;U|Y)} 
\vspace{1mm} \\
+\bar{\mu}\ByShun{D(q_{X|UY}||p_{X|Y}|q_{UY})}
+\mu {\rm E}_q d(X,Z)]
\ea
\\
& &-\frac{1}{2}\rho(p_{XY})\lambda^2
\\
&\geq& 
\ba[t]{l}
 \bar{\alpha}[ D(q_X||p_X)
+ D(q_{Y|XU}||p_{Y|X}|q_{XU})
\vspace{1mm}\\
\hspace*{2mm}+I_q(X;Z|UY)]
\vspace{1mm}\\
+ 4\alpha[\bar{\mu}\ByShun{I_q(X;U|Y)} 
+\mu {\rm E}_q d(X,Z)]
\ea
\\
& &-\frac{1}{2}\rho(p_{XY})\lambda^2.
\eeqno
Specifically, we have 
$$
\Omega^{(\alpha, \mu, \lambda)}(p_{XY}) 
\geq \lambda \tilde{R}^{(\alpha,\mu)}(p_{XY})
-\frac{1}{2}\rho(p_{XY})\lambda^2.
$$

\item[d)] 
For any $\delta>0$, there exists a positive number
$
\nu
=\nu(\delta,d_{\max},|{\cal X}|,|{\cal Y}|,|{\cal Z}|)\in (0,1] 
$
such that 
for every $\tau \in (0,\nu],$
the condition 
$(R+\tau, {\Dist} +\tau) \notin {\cal R}(p_{XY})$
implies 
$$
F(R,{\Dist}|p_{XY})\geq \frac{\rho(p_{XY})}{2}\cdot g^2
\left(\ts \frac{\tau^{3+\delta}}{\rho(p_{XY})}\right)>0,
$$
where $g$ is the inverse function of $\vartheta(a) \defeq (1/2)a+a^2, a>0$.
\end{itemize}

\end{pr}

Proof of Property \ref{pr:pro1} part a) is {\GorF} in Appendix \ref{sub:ApdaAAA}.
\newcommand{\ApdaAAAb}{

Next we give a proof of Property \ref{pr:pro1} part a). 

{\it Proof of Property \ref{pr:pro1} part a):} 
We bound  the cardinality $|{\cal U}|$ of ${U}$ 
to show that the bound 
$|{\cal U}|\leq |{\cal X}|$$|{\cal Y}|$$|{\cal Z}|$
is sufficient to describe 
${\Omega}{(\alpha, \mu,\lambda)}$ 
$(p_{XY})$. Observe that 
\beqa
& &q_{XYZ}(x,y,z)
=\sum_{u \in {\cal U}}q_{U}(u)
q_{XYZ|{U}}(x,y,z|u),
\label{eqn:asdfs}
\\
& &\exp
\left\{-\Omega^{(\alpha,\mu,\lambda)}(q|p_{XY})
\right\}
\nonumber\\
& &\qquad\quad
=\sum_{u\in {\cal U}}q_{{U}}(u)
\Pi^{(\alpha,\mu,\lambda)}
(q_{{XYZ}|{U}}(\cdot,\cdot,\cdot|u)),
\label{eqn:aqqqas}
\eeqa
where 
\beqno
& &\Pi^{(\alpha,\mu,\lambda)}
(q_{XYZ|{U}}(\cdot,\cdot,\cdot|u)) 
\\
& \defeq &
\sum_{\scs (x,y,z) 
\atop{  \in{\cal X}
         \times{\cal Y}
         \times{\cal Z}
     }
}
\hspace*{-3mm}
q_{XYZ|U}(x,y,z|u)
%
\exp\left\{-\lambda \omega^{(\alpha,\mu)}_{{\pANDq}}
(x,y,z|u) \right\}.
\eeqno
For each $u\in {\cal U}$, 
$
\Pi^{(\alpha,\mu)}(q_{XYZ|U}(\cdot,\cdot,\cdot|u))
$ is a 
continuous function of $q_{XYZ|U}(\cdot,\cdot,\cdot|u)$.
Then by the support lemma, 
$$
|{\cal U}|\leq |{\cal X}||{\cal Y}||{\cal Z}|-1 +1 = 
|{\cal X}||{\cal Y}||{\cal Z}|
$$
is sufficient to express $|{\cal X}||{\cal Y}|$ $|{\cal Z}|-1$ 
values of (\ref{eqn:asdfs}) and one value of (\ref{eqn:aqqqas}).
\hfill \IEEEQED
}
%
%
Proof of Property \ref{pr:pro1} parts b), c) and d) are 
{\GorF} in Appendix \ref{sub:ApdaAAC}. 
\newcommand{\ApdaAAC}{
\subsection{
Proof of Property \ref{pr:pro1} parts b), c), and d) 
}
\label{sub:ApdaAAC}


In this appendix we prove Property \ref{pr:pro1}
parts b), c), and d). 

{\it Proof of Property \ref{pr:pro1} parts b), c), and d):} \ We 
first prove parts b) and c). For simplicity of notations, set
\beqno
& & 
\underline{a} \defeq (u,x,y,z), 
\underline{A}\defeq (U,X,Y,Z),
\underline{\cal A} \defeq 
 {\cal U} \times 
 {\cal X} \times 
 {\cal Y} \times 
 {\cal Z},
\\
& & \omega^{(\mu)}_{q_{XYZ|U} }(x,y,z|u) 
\defeq g(\underline{a}),
\\
& & 
\OMeGa
\defeq \xi(\lambda).
\eeqno
Then we have 
$$
\xi(\lambda)=-\log
\left
[\sum_{\underline{a}\in \underline{\cal A} }p_{\underline{A}}(\underline{a})
{\rm e}^{-\lambda g(\underline{a})}\right].
$$
We set 
$$
q^{(\lambda)}(u,x,y,z)\defeq p_{\underline{A}}^{(\lambda)}(\underline{a}).
$$
Then 
$
p_{\underline{A}}^{(\lambda)}(\underline{a}), \underline{a}\in {\cal A}
$
has the following form:
$$
p_{\underline{A}}^{(\lambda)}(\underline{a})
={\rm e}^{\xi(\lambda)}
p_{\underline{A}}(\underline{a})
{\rm e}^{-\lambda{g}(\underline{a})}.
$$
By simple computations we have 
\beqa
\xi^{\prime}(\lambda)&=&
{\rm e}^{\xi(\lambda)}
\left[
\sum_{\underline{a}\in \underline{\cal A}} 
p_{\underline{A}}(\underline{a})
{g}(\underline{a}) 
{\rm e}^{-\lambda{g}(\underline{a})}\right]
\nonumber\\
&=& \sum_{\underline{a}\in \underline{\cal A}} 
p_{\underline{A}}^{(\lambda)}(\underline{a})
{g}(\underline{a}),
\nonumber
\eeqa
\beqa
\hspace*{-5mm}&&\xi^{\prime\prime}(\lambda)=-{\rm e}^{2\xi(\lambda)}
\nonumber\\
\hspace*{-5mm}&&\times 
\left[
\sum_{\underline{a}, \underline{b}\in \underline{\cal A}}
 p_{\underline{A}}(\underline{a})p_{\underline{A}}(\underline{b})
\frac{\left\{{g}(\underline{a})-{g}(\underline{b})\right\}^2}{2}
{\rm e}^{-\lambda\left\{{g}(\underline{a})
+{g}(\underline{b})\right\}}\right]
\nonumber\\
\hspace*{-5mm}&&=-
\sum_{\underline{a}, \underline{b}\in \underline{\cal A}}
p_{\underline{A}}^{(\lambda)}(\underline{a})
p_{\underline{A}}^{(\lambda)}(\underline{b})
\frac{\left\{{g}(\underline{a})-{g}(\underline{b})\right\}^2}{2}
\nonumber\\
\hspace*{-5mm}&&=-
\sum_{\underline{a}\in \underline{\cal A}}
p_{\underline{A}}^{(\lambda)}(\underline{a})
{g}^2(\underline{a})
+\left[
\sum_{\underline{a}\in \underline{\cal A}} 
p_{\underline{A}}^{(\lambda)}(\underline{a})
{g}(\underline{a})\right]^2.
\nonumber
\eeqa
By the Taylor expansion of 
$\Omega^{(\alpha, \mu, \lambda)}(q|p_{XY})$
with respect to $\lambda$ around $\lambda=0$,
we have 
\beqno
& &\Omega^{(\alpha, \mu, \lambda)}(q | p_{XY} )
\\
&=&\xi(\lambda)=\xi(0)+ \xi^{\prime}(0)\lambda
+\frac{1}{2}\xi^{\prime\prime}(0)(\tau\lambda)^2
\\
&=&
{\rm E}_{q}
\left[
\omega^{(\alpha, \mu)}_{{\pANDq}}(X,Y,Z|U)
\right]
\\
& &-\frac{1}{2}(\tau \lambda)^2 {\rm Var}_{q}
\left[
\omega^{(\alpha, \mu)}_{{\pANDq}}(X,Y,Z|U)\right]
\eeqno
for some $\tau \in [0,1]$. Then by the definition 
of $\rho$, we have  
\beqno
& &\Omega^{(\alpha, \mu,\lambda)}(q|p_{XY}) 
\\
&\geq& \lambda {\rm E}_{q}
\left[\omega^{(\alpha, \mu)}_{{\pANDq}}(X,Y,Z|U)\right]
-\frac{1}{2}\rho\lambda^2.
\eeqno
The second inequality is obvious from the first inequality. 
We finally prove the part d). 
By the hyperplane expression 
${\cal R}_{\rm sh}(p_{XY})$ of 
${\cal R}(p_{XY})$ 
stated Property \ref{pr:pro0z} part b)
we have that when 
$(R+\tau, \Dist+\tau) \notin {\cal R}(p_{XY})$, 
we have 
\beq
\bar{\mu}^*(R+\tau)+\mu^*(\Dist+\tau)
<
R^{(\mu^*)}(p_{XY})
\label{eqn:Zafgg}
\eeq
for some $\mu^*\in [0,1]$. 
Let $\alpha_0=\alpha_0(d_{\max},|{\cal X}|)$ 
be the quantity defined by (\ref{eqn:AlpaConeCtwo}) 
in Property \ref{pr:pro0z} part b). 
Then from (\ref{eqn:Zafgg}), 
we have that for any $\alpha \in (0, \alpha_0]$, 
\beqa 
& &\bar{\mu}^* R +\mu^*{\Dist} +\tau \leq  R^{(\mu^*)}(p_{XY})
\nonumber\\
&\MLeq{a}& 
\frac{1}{4\alpha}\tilde{R}^{(\alpha, \mu^*)}(p_{XY})
  +c_1\sqrt{\frac{\alpha}{\bar{\alpha}}}
\log\left(c_2\frac{\bar{\alpha}}{\alpha}\right).
\label{eqn:Sdxx}
\eeqa 
Step (a) follows from the first inequality of (\ref{eqn:PropEqA}) 
in Property \ref{pr:pro0z} part b). 
Fix any $\delta>0$. We choose $\alpha=\tau^{2+\delta}$.
For this choice of $\alpha$, we have (\ref{eqn:Sdxx}) 
for any $\tau \in (0,\alpha_0^{\frac{1}{2+\delta}}]$ 
and 
\beqno
& &
c_1 \cdot \sqrt{\frac{\alpha}{\bar{\alpha}}}
\log\left(c_2\frac{\bar{\alpha}}{\alpha}\right)
\times \frac{1}{\tau}
\\
&=&c_1\frac{\tau^{\frac{\delta}{2}}}
{\sqrt{1-\tau^{2+\delta}}}
\log\left(c_2
\cdot\frac{1-\tau^{2+\delta}}{\tau^{2+\delta}}\right)
\to 0\mbox{ as }\tau \to 0. 
\eeqno
Hence there exists a positive 
$\nu=\nu( \delta, d_{\max}, |{\cal X}|,|{\cal Y}|,|{\cal Z}|)$ 
with $\nu\leq \alpha_0^{\frac{1}{2+\delta}}$ such that for 
$
\tau \in (0,\nu] \subseteq (0,\alpha_0^{\frac{1}{2+\delta}}],
$
\beq
c_1 \cdot \sqrt{\frac{\alpha}{\bar{\alpha}}}
\log\left(c_2\frac{\bar{\alpha}}{\alpha}\right)
\leq \frac{\tau}{2}.
\eeq
The above inequality together with (\ref{eqn:Sdxx}) yields 
that for $\tau\in (0,\nu]$,
\beq
\bar{\mu}^{\ast} R+\mu^*{\Dist}+\frac{1}{2} \tau \leq 
\frac{1}{4\tau^{2+\delta}}
\tilde{R}^{(\tau^{2+\delta},\mu^*)}(p_{XY}).
\label{eqn:Xddcc}
\eeq
Then for each $\tau\in (0, \nu ],$
we have the following chain of inequalities: 
\beqno
& & F(R,{\Dist}|p_{XY})
\\
&\geq &\sup_{\lambda>0}
F^{(\tau^{2+\delta},\mu^{*},\lambda)}
(\bar{\mu}^{\ast} R+\mu^{*}{\Dist}|p_{XY})
\\
&=&
\sup_{\lambda >0 }
\frac{\Omega^{(\tau^{2+\delta}, \mu^*,\lambda)}(p_{XY})
  -4\lambda \tau^{2+\delta}(\bar{\mu}^{\ast} R+ \mu^*{\Dist})}
{1+4(1-\mu^*\tau^{2+\delta})\lambda}
\\
&\MGeq{a} & 
\sup_{\lambda>0}
\frac{1}{1+4\lambda}
\left\{ 
-\frac{1}{2}\rho \lambda^2
+\lambda \tilde{R}^{(\tau^{2+\delta},\mu^*)}(p_{XY})\right.
\\
& &\qquad \qquad -4\lambda\tau^{2+\delta}(\bar{\mu}^{\ast} 
R+\mu^*{\Dist})\Bigr\}
\\
&\MGeq{b}& 
\sup_{\lambda>0}
\frac{1}{1+4\lambda}
\left\{
-\frac{1}{2}\rho \lambda^2+ 2\tau^{3+\delta}\lambda
\right\}
\MEq{c}
\frac{\rho}{2}g^2\left( \frac{\tau^{3+\delta}}{\rho}\right).
\eeqno
Step (a) follows from Property \ref{pr:pro1} part b).
Step (b) follows from (\ref{eqn:Xddcc}).
Step (c) follows from an elementary computation.
This completes the proof of Property \ref{pr:pro1} part d). 
\hfill\IEEEQED
}
Our main result is the following. 

\begin{Th}\label{Th:main} 
For any $R,\Dist>0$, any $p_{XY}$, and 
for any $(\varphi^{(n)},$ $\psi^{(n)})$ satisfying 
$$
\frac{1}{n}\log {M_n}\leq R,  
$$
we have 
\beq
{\rm P}_{\rm c}^{(n)}(\varphi^{(n)},\psi^{(n)};{\Dist})
\leq 5\exp \left\{-n F(R,\Dist|p_{XY})\right\}.
\label{eqn:mainIeq}
\eeq
\end{Th}

It follows from Theorem \ref{Th:main} and 
Property \ref{pr:pro1} part d) that if $(R,\Dist)$ is outside the rate distortion 
region, then the error probability of decoding goes to one 
exponentially and its exponent is not below 
$F(R,\Dist|p_{XY})$. 

It immediately follows from Theorem \ref{Th:main} 
that we have the following corollary. 
\begin{co}\label{co:mainCo} 
For any $R,\Dist>0$ and any $p_{XY}$, we have 
\beqa 
& &G(R,\Dist|p_{XY})\geq F(R,\Dist|p_{XY}).
\eeqa
Furthermore for any $p_{XY}$, we have
\beqa
& &{\cal G}(p_{XY})
\subseteq  \overline{\cal G}(p_{XY})
\nonumber\\
& &\qquad\qquad\defeq \left\{(R,\Dist,G): G\geq F(R,\Dist|p_{XY})\right\}.
\eeqa
\end{co}

Proof of Theorem \ref{Th:main} will be given in the next section. The 
exponent function in the case of ${\Dist}=0$ can be obtained as a 
corollary of the result of Oohama and Han \cite{OhHan94} for the 
separate source coding problem of correlated sources \cite{sw}. The 
techniques used by them is a method of types \cite{ckBook81}, which is 
not useful for proving Theorem \ref{Th:main}. In fact when we use this 
method, it is very hard to extract a condition related to the Markov 
chain condition $U \leftrightarrow X \leftrightarrow Y$, which the 
auxiliary random variable $U\in {\cal U}$ must satisfy when $(R,\Dist)$ 
is on the boundary of the set ${\cal R}(p_{XY})$. Some novel techniques 
based on the information spectrum method introduced by Han \cite{han} 
are necessary to prove this theorem.

From Theorem \ref{Th:main} and Property \ref{pr:pro1} part d), 
we obtain an explicit outer bound of 
${\cal R}_{\rm WZ}(\varepsilon |p_{XY})$ 
with an asymptotically vanishing deviation 
from ${\cal R}_{\rm WZ}(p_{XY})$ 
$={\cal R}(p_{XY})$. 
The strong converse theorem 
{ } immediately follows from this corollary.
From Theorem \ref{Th:main} and Property \ref{pr:pro1} part d) 
we have the following corollary.
\begin{co}\label{co:StConv}
For each fixed $\varepsilon$ $ \in (0,1)$ 
and for any $\delta>0$, there exists a positive integer $n_0$ with 
$$
n_0=n_0(\varepsilon,
\delta,d_{\max},|{\cal X}|,|{\cal Y}|,|{\cal Z}|,\rho(p_{XY}))
$$
such that for $n\geq n_0$, we have
\beqno
&&{\cal R}_{\rm WZ}(\varepsilon |p_{XY})
\subseteq 
{\cal R}(p_{XY})-\kappa_n(1,1), 
\eeqno
where
\beqa
\kappa_n
&\defeq&
\left\{
\sqrt{\ts
\frac{\rho(p_{XY})}{2n}
\log\left(\frac{5}{1-\varepsilon}\right)}
+\ts\frac{2}{n}
\log\left(\frac{5}{1-\varepsilon}\right)
\right\}^{\frac{1}{3+\delta}}.
\nonumber
\eeqa
It immediately follows from the above result that
for each fixed $\varepsilon\in (0,1)$, we have 
\beqno
& &{\cal R}_{\rm WZ}(\varepsilon | p_{XY})
  ={\cal R}_{\rm WZ}(p_{XY})
  ={\cal R}(p_{XY}).
\eeqno
\end{co}

Proof of this corollary {\GorFb}.
\newcommand{\ProofCor}{  
{\it Proof of Corollary \ref{co:StConv}:} 
To prove this corollary we use the following 
expression of ${\cal R}_{\rm WZ}(\varepsilon| p_{XY)}$ stated in 
Property \ref{pr:pro0a} part b): 
\beq
{\cal R}_{\rm WZ}(\varepsilon|p_{XY})
={\rm cl}\left(
\bigcup_{m\geq 1}
\bigcap_{n \geq m}{\cal R}_{\rm WZ}(n,\varepsilon | p_{XY})
\right).
\label{eqn:Zassd0}
\eeq 
We assume that 
\beq
(R,\Dist)
\in 
\bigcup_{m\geq 1}
\bigcap_{n \geq m}{\cal R}_{\rm WZ}(n,\varepsilon | p_{XY}).
\label{eqn:Zass}
\eeq 
Then there exists a positive integer 
$m_0=m_0(\varepsilon)$ and some 
$\{(\varphi^{(n)},$ $\psi^{(n)})$ $\}_{n\geq m_0}$
such that for $n\geq m_0(\varepsilon)$, we have
$$
\frac{1}{n}\log ||\varphi^{(n)}||\leq R, 
\quad {\rm P}_{\rm e}^{(n)}(\varphi^{(n)},\psi^{(n)};{\Dist})
\leq \varepsilon.
$$
Then by Theorem \ref{Th:main}, we have
\beqa
1-\varepsilon&\leq& 
{\rm P}_{\rm c}^{(n)}(\varphi^{(n)},\psi^{(n)};{\Dist})
\nonumber\\
&\leq& 5\exp \left\{-n F(R,\Dist|p_{XY})\right\}
\label{eqn:Zsddd}
\eeqa
for any $n\geq m_0(\varepsilon)$. Fix any $\delta>0$.
We take a positive number
$
\nu
=\nu(\delta,d_{\max},|{\cal X}|,|{\cal Y}|,|{\cal Z}|) \in (0,1] 
$
appearing in Property \ref{pr:pro1} part d) and set
\beqa
\kappa_n&=&\left\{
\rho(p_{XY}) \vartheta\left(
\sqrt{\ts \frac{2}{n\rho(p_{XY})} \log\left(\frac{5}{1-\varepsilon}\right)} 
\right)
\right\}^{\frac{1}{3+\delta}}
\label{eqn:zdd}\\
&=&
\left\{
\sqrt{\ts
\frac{\rho(p_{XY})}{2n}
\log\left(\frac{5}{1-\varepsilon}\right)}
+\ts\frac{2}{n}
\log\left(\frac{5}{1-\varepsilon}\right)
\right\}^{\frac{1}{3+\delta}}.
\nonumber
\eeqa
Since $g$ is an inverse function of $\vartheta$, (\ref{eqn:zdd}) 
is equivalent to 
\beq
g\left(\ts \frac{\kappa_{n}^{3+\delta}}{\rho(p_{XY})}\right)
=\sqrt{\ts \frac{2}{n\rho(p_{XY})} \log\left(\frac{5}{1-\varepsilon}\right)}.
\label{eqn:zddQ}
\eeq
We take a sufficiently large positive integer $m_1$ so that
we have 
$\kappa_n < \nu$
for $n\geq m_1$. Set $n_0=\max\{m_0,$ $m_1\}$. 
We claim that for $n\geq n_0$, we have 
$(R+\kappa_n, {\Dist}+\kappa_n)$ $\in$ ${\cal R}(p_{XY})$. 
To prove this claim we suppose that  
$(R+\kappa_{n^*}, {\Dist}+\kappa_{n^*})$
does not belong to ${\cal R}(p_{XY})$ for some $n^*\geq n_0$. 
Since ${\cal R}(p_{XY})$ is a closed set, there exists 
a positive number $\tau> \kappa_{n^*} $ sufficiently 
close to $\kappa_{n^*}$ such that 
\beqno
&&\kappa_{n^*}<\tau \leq \nu,
\\
&&(R+\tau, {\Dist} +\tau)\notin {\cal R}(p_{XY}).
\eeqno
Then we have the following chain of inequalities:
\beqa
& &5\exp\left[-n^* F(R,{\Dist}|p_{XY}) \right]
\nonumber\\
&\MLeq{a}&5\exp\left[-
\frac{n^*\rho(p_{XY})}{2}\cdot 
g^2\left(\ts \frac{\tau^{3+\delta}}{\rho(p_{XY})}\right)
\right]
\nonumber\\
&\ML{b}&
5\exp\left[-
\frac{n^*\rho(p_{XY})}{2}
\cdot g^2\left(\ts \frac{\kappa_{n^*}^{3+\delta}}{\rho(p_{XY})}\right)
\right]
\nonumber\\
&\MEq{c}&
5\exp\left[
-\frac{n^*\rho(p_{XY})}{2}\frac{2}{n^*\rho(p_{XY})}
\log\left(\frac{5}{1-\vep}\right)
\right]
\nonumber\\
&=&1-\vep.
\label{eqn:Dssd}
\eeqa 
Step (a) follows from Property \ref{pr:pro1} part d).
Step (b) follows from $\kappa_{n^*} < \tau.$
Step (c) follows from (\ref{eqn:zddQ}).
The bound (\ref{eqn:Dssd}) contradicts (\ref{eqn:Zsddd}). 
Hence we have $(R+\kappa_n,\Dist+$$\kappa_n)$
$\in$${\cal R}(p_{XY})$ or equivalent to
$$
(R,\Dist)\in {\cal R}(p_{XY})-\kappa_n(1,1)
$$
for $n\geq n_0$. Recalling the first assumption (\ref{eqn:Zass}) 
on $(R,{\Dist})$, we obtain
\beq
\bigcup_{m\geq 1}
\bigcap_{n \geq m}{\cal R}_{\rm WZ}(n,\varepsilon | p_{XY})
\subseteq
{\cal R}(p_{XY})-\kappa_n(1,1).
\label{eqn:SddXX}
\eeq 
Taking the closure of both sides of 
(\ref{eqn:SddXX}), using (\ref{eqn:Zassd0}), 
and considering 
that ${\cal R}(p_{XY})$$-\kappa_n(1,1)$ is a closed set, 
we have that for $n\geq n_0$,
\beqno
& &{\cal R}_{\rm WZ}(\varepsilon | p_{XY})
\\
&=&{\rm cl}\left(\bigcup_{m\geq 1}
\bigcap_{n \geq m}{\cal R}_{\rm WZ}(n,\varepsilon | p_{XY})
\right)
\subseteq
{\cal R}(p_{XY})-\kappa_n(1,1),
\eeqno
completing the proof. 
\hfill\IEEEQED
}
The direct part of coding theorem, i.e., the inclusion of 
${\cal R}(p_{XY}) $ $\subseteq$ ${\cal R}_{\rm WZ}(\varepsilon|p_{XY})$
was established by Csisz\'ar and K\"orner \cite{ckBook81}. They proved 
a weak converse theorem to obtain the inclusion ${\cal R}_{\rm WZ}(p_{XY})$ 
$\subseteq {\cal R}(p_{XY})$. Until now we have had no result on the 
strong converse theorem. The above corollary stating the strong 
converse theorem for the Wyner-Ziv source coding problem implies that a 
long standing open problem since Csisz\'ar and K\"orner \cite{ckBook81} 
has been resolved. 

\section{Proof of the Main Result}

In this section we prove Theorem \ref{Th:main}. 
We first present a lemma
which upper bounds the correct \ByShun{probability} of decoding by the 
information spectrum quantities. 
We set 
$$
S_n\defeq\varphi^{(n)}(X^n), 
Z^n\defeq\psi_n(\varphi^{(n)}(X^n),Y^n).
$$ 
It is obvious that 
$$
S_n \markov X^n \markov Y^n,
X^n \markov (S_n,Y^n) \markov Z^n.
$$
Then we have the following.
\begin{lm}\label{lm:Ohzzz} For any $\eta>0$ and for 
any $(\varphi^{(n)}$, $\psi^{(n)})$ satisfying 
$$
\frac{1}{n} \log M_n \leq R,
$$
we have 
\beqa
\lefteqn{\hspace*{-8mm}{\rm P}_{\rm c}^{(n)}(\varphi^{(n)}, \psi^{(n)};{\Dist})
\leq p_{S_nX^nY^nZ^n}\biggl\{
}
\nonumber\\
\eta &\geq& \frac{1}{n}\log \frac{q^{\One}_{{X}^n}(X^n)}{p_{X^n}(X^n)},
\label{eqn:asppa}\\
\eta &\geq& \frac{1}{n}\log \frac{q^{\Two}_{Y^n|S_nX^n}(Y^n|S_n,X^n)}
                                 {p_{Y^n|X^n}(Y^n|X^n)},
\label{eqn:asppb}\\
\eta &\geq& \frac{1}{n}\log 
          \frac{q^{\Thr}_{X^n|S_nY^nZ^n}(X^n|S_n,Y^n,Z^n)}
               {p_{X^n|S_nY^n}(X^n|S_n,Y^n)},
\label{eqn:asppc}\\
R+\eta&\geq&\frac{1}{n}\log
\frac{q^{\Fou}_{X^n|S_nY^n}(X^n|S_n,Y^n)}
      {p_{X^n|Y^n}(X^n|Y^n)},
\label{eqn:asppd}\\
{\Dist} &\geq & \left.
\frac{1}{n}\log \exp \left\{ d(X^n,Z^n) \right\}
\right\}
+4{\rm e}^{-n\eta}. 
\label{eqn:azsada}
\eeqa
The probability distribution and stochastic matrices 
appearing in the right members of 
(\ref{eqn:azsada}) have a property that we can select them 
arbitrary. In (\ref{eqn:asppa}), we can choose any probability 
distribution 
$q^{\One}_{{X}^n}$ on ${\cal X}^n$. In (\ref{eqn:asppb}), we can choose 
any stochastic matrix $q^{\Two}_{Y^n|S_nX^n}:$ ${\cal M}_n\times$ 
${\cal X}^n$ $\to {\cal Y}^n$. In (\ref{eqn:asppc}), 
we can choose any stochastic matrix 
$q^{\Thr}_{X^n|S_nY^nZ^n}:$ 
${\cal M}_n\times$ ${\cal Y}^n$ $\times$ ${\cal Z}^n$ $\to {\cal X}^n$.
In (\ref{eqn:asppd}), we can choose any stochastic matrix 
$q^{\Fou}_{X^n|S_nY^n}:$ ${\cal M}_n$ $\times$ ${\cal Y}^n$ $\to {\cal X}^n$.
\end{lm}

Proof of this lemma is {\GorF} in Appendix \ref{sub:Apda}.
\newcommand{\Apda}{
\subsection{
Proof of Lemma \ref{lm:Ohzzz}
}\label{sub:Apda}

To prove Lemma \ref{lm:Ohzzz}, we prepare a lemma. Set
\beqno
\tilde{\cal A}_n
&\defeq &
\left\{x^{n}: 
\frac{1}{n}\log 
\frac {p_{X^n}(x^n)}{q^{\One}_{{X}^n}(x^n)}
\geq -\eta
\right\},
\\
{\cal A}_n 
&\defeq & 
\tilde{\cal A}_n \times {\cal M}_n
\times{\cal Y}^n \times {\cal Z}^n,
\\
{\cal A}_n^{\rm c}
& \defeq & \tilde{\cal A}_n^{\rm c}
  \times {\cal M}_n\times {\cal Y}^n \times {\cal Z}^n,
\\
\tilde{\cal B}_n
&\defeq &
\left\{(s,x^n,y^n): 
\frac{1}{n}\log 
\frac {p_{Y^n|X^n}(y^n |x^n)}{q^{\Two}_{Y^n|X^nS_n}(y^n|x^n,s)}
\geq -\eta
\right\},
\\
{\cal B}_n 
  &\defeq& \tilde{\cal B}_n \times {\cal Z}^n,
  {\cal B}_n^{\rm c}
  \defeq \tilde{\cal B}_n^{\rm c}\times {\cal Z}^n,
\\
{\cal C}_n 
&\defeq &\Bigl\{(s,x^n,y^n,z^n):
\\
& &
\left.\frac{1}{n}\log 
\frac{p_{X^n|S_nY^n}(x^n|s,y^n)}
     {q^{\Thr}_{X^n|S_nY^nZ^n}(x^n|s,y^n,z^n)}\geq -\eta
\right\},
\\
\tilde{\cal D}_n
&\defeq& \{(s,x^n,y^n): s=\varphi^{(n)}({x^n}),
\\ 
&&\quad q^{\Fou}_{X^n|S_nY^n}(x^n|s,y^n)\leq M_n {\rm e}^{n\eta}
        p_{X^n|Y^n}(x^n|y^n) \}, 
\\
{\cal D}_n&\defeq& \tilde{\cal D}_n\times {\cal Z}^n,
  {\cal D}_n^{\rm c}
  \defeq \tilde{\cal D}_n^{\rm c}\times {\cal Z}^n.
\eeqno
Then we have the following lemma. 
\begin{lm}\label{lm:zzxa}{
\beqno
& &
p_{S_nX^nY^nZ^n}
\left({\cal A}_n^{\rm c}\right)\leq {\rm e}^{-n\eta}, 
p_{S_nX^nY^nZ^n}
\left({\cal B}_n^{\rm c}\right)\leq {\rm e}^{-n\eta}, 
\\
& &p_{S_nX^nY^nZ^n}
\left({\cal C}_n^{\rm c} \right)\leq {\rm e}^{-n\eta},
p_{S_nX^nY^nZ^n}
\left( {\cal D}_n^{\rm c} \right)
\leq {\rm e}^{-n\eta}.
\eeqno
}
\end{lm}

{\it Proof:} We first prove the first inequality. 
We have the following chain of inequalities:
\beqno
& &p_{S_nX^nY^nZ^n}
({\cal A}_n^{\rm c})
= p_{X^n}
(\tilde{\cal A}_n^{\rm c})
=\sum_{x^n\in \tilde{\cal A}_n^{\rm c} } p_{X_n}(x^n)
\\
&\MLeq{a}&\sum_{x^n\in \tilde{\cal A}_n^{\rm c}}
{\rm e}^{-n\eta}q^{\One}_{{X}^n }(x^n)
\leq {\rm e}^{-n\eta} \sum_{x^n}q^{\One}_{{X}^n }(x^n)={\rm e}^{-n\eta}.
\eeqno
Step (a) follows from the definition of ${\cal A}_n$. 
We next prove the second inequality. We have the following 
chain of inequalities:
\beqno
& &p_{S_nX^nY^nZ^n}
({\cal B}_n^{\rm c})
= p_{S_nX^nY^n}
(\tilde{\cal B}_n^{\rm c})
\\
&\MEq{a}&
\sum_{(s,x^n,y^n) \in \tilde{\cal B}_n^{\rm c} } 
p_{S_nX^n}(s,x^n)p_{Y^n|X^n}(y^n|x^n)
\\
&\MLeq{b}&
\sum_{(s,x^n,y^n) \in \tilde{\cal B}_n^{\rm c}}
{\rm e}^{-n\eta}p_{S_nX^n}(s,x^n)q^{\Two}_{Y^n|S_nX^n}(y^n|s,x^n)
\\
&\leq&{\rm e}^{-n\eta}
\sum_{s,x^n,y^n} p_{S_nX^n}(s,x^n)q^{\Two}_{Y^n|S_nX^n}(y^n|s,x^n)
={\rm e}^{-n\eta}.
\eeqno
Step (a) follows from the Markov chain 
$S_n \markov$  $X^n \markov$ $Y^n$. 
Step (b) follows from the definition of ${\cal B}_n$.   
On the third inequality we have the following 
chain of inequalities:
\beqno
& &p_{S_nX^nY^nZ^n}({\cal C}_n^{\rm c})
\\
&\MEq{a}&
\sum_{(s,x^n,y^n,z^n) \in {\cal C}_n^{\rm c}} 
p_{X^n|S_nY^n}(x^n|s,y^n)p_{S_nY^nZ^n}(s,y^n,z^n)
\\
&\MLeq{b}&
\sum_{(s,x^n,y^n,z^n) \in {\cal C}_n^{\rm c}}
{\rm e}^{-n\eta}q^{\Thr}_{X^n|S_nY^nZ^n}(x^n|s,y^n,z^n)
\\
& & \qquad \times p_{S_nY^nZ^n}(s,y^n,z^n)
\\
&\leq&{\rm e}^{-n\eta}
\sum_{s,x^n,y^n,z^n}
q^{\Thr}_{X^n|S_nY^nZ^n}(x^n|s,y^n,z^n)
\\
& &\qquad \times p_{S_nY^nZ^n}(s,y^n,z^n)
={\rm e}^{-n\eta}.
\eeqno
Step (a) follows from the Markov chain 
$X^n \markov$  $S_nY^n \markov$ $Z^n$. 
Step (b) follows from the definition of ${\cal C}_n$.   
We finally prove the fourth inequality. We have 
the following chain of inequalities:
\beqno
& &p_{S_nX^nY^nZ^n}
   ({\cal D}_n^{\rm c})
   =p_{S_nX^nY^n}(\tilde{\cal D}_n^{\rm c})
\\
&=&\sum_{s\in{\cal M}_n}
\sum_{\scs (x^n,y^n):\varphi^{(n)}({x^n})=s,
   \atop{\scs  p_{X^n|Y^n}(x^n|y^n)
     \atop{\scs  \leq (1/M_n){\rm e}^{-n\eta}\qquad
        \atop{\scs \times q^{\Fou}_{X^n|S_n,Y^n}(x^n|s,y^n)
        }
     }
  }
}
p_{X^n|Y^n}(x^n|y^n)p_{Y^n}(y^n)
\\
&\leq &
\frac{{\rm e}^{-n\eta}}{M_n}
\sum_{s\in{\cal M}_n}
\sum_{\scs (x^n,y^n): \varphi^{(n)}({x^n})=s,
\atop{\scs p_{X^n|Y^n}(x^n|y^n)
       \atop{\scs \leq (1/M_n){\rm e}^{-n\eta}\qquad
           \atop{ \scs \times q^{\Fou}_{X^n|S_nY^n}(x^n|s,y^n)
            }
       }
     }
}
\hspace*{-7mm}
q^{\Fou}_{X^n|S_nY^n}(x^n|s,y^n)p_{Y^n}(y^n)
\\
&\leq &
\frac{{\rm e}^{-n\eta}}{M_n}
\sum_{s\in{\cal M}_n}\sum_{\scs x^n,y^n}
q^{\Fou}_{X^n|S_nY^n}(x^n|s,y^n)p_{Y^n}(y^n)
={\rm e}^{-n\eta}. 
\eeqno
\hfill\IEEEQED

{\it Proof of Lemma \ref{lm:Ohzzz}:} 
We set 
$$
{\cal E}_n\defeq \left\{(s,x^n,y^n,z^n): 
\frac{1}{n} d(X^n,Z^n)\leq {\Dist} \right\}. 
$$
Set $R^{(n)}\defeq (1/n)\log M_n$. By definition 
we have
\beqa
\lefteqn{
\hspace*{-20mm}
p_{S_nX^nY^nZ^n}
\left(
{\cal A}_n\cap {\cal B}_n\cap {\cal C}_n
\cap {\cal D}_n \cap {\cal E}_n
\right)}
\nonumber\\
= p_{S_nX^nY^nZ^n}\biggl\{
\eta &\geq& \frac{1}{n}
\log \frac{q^{\One}_{{X}^n}(X^n)}{p_{X^n}(X^n)},
\nonumber\\
\eta &\geq& \frac{1}{n}
\log \frac{q^{\Two}_{Y^n|X^nS}(Y^n|X^nS)}{p_{Y^n|X^n}(Y^n|X^n)},
\nonumber\\
\eta &\geq& \frac{1}{n}\log 
          \frac{q^{\Thr}_{X^n|S_nY^nZ^n}(X^n|S_nY^nZ^n)}
               {p_{X^n|S_nY^n}(X^n|S_nY^n)},
\nonumber\\
R^{(n)}+\eta&\geq&
\frac{1}{n}\log
\frac{q^{\Fou}_{X^n|S_nY^n}(X^n|S_nY^n)}{p_{X^n|Y^n}(X^n|Y^n)},
\nonumber\\
{\Dist} &\geq & 
\left.
\frac{1}{n}\log \exp \left\{ d(X^n,Z^n) \right\}
\right\}.
\label{eqn:azsadad}
\eeqa
Then for any $(\varphi^{(n)}$, $\psi^{(n)})$ satisfying 
$$
R^{(n)}=\frac{1}{n}\log {M_n}\leq R,
$$
we have 
\beqa
\lefteqn{
\hspace*{-20mm}
p_{S_nX^nY^nZ^n}
\left(
{\cal A}_n\cap {\cal B}_n\cap {\cal C}_n
\cap {\cal D}_n \cap {\cal E}_n
\right)}
\nonumber\\
\leq p_{S_nX^nY^nZ^n}\biggl\{
\eta &\geq& \frac{1}{n}\log \frac{q^{\One}_{{X}^n}(X^n)}{p_{X^n}(X^n)},
\nonumber\\
\eta &\geq& \frac{1}{n}\log \frac{q^{\Two}_{Y^n|X^nS}(Y^n|X^nS)}{p_{Y^n|X^n}(Y^n|X^n)},
\nonumber\\
\eta &\geq& \frac{1}{n}\log 
          \frac{q^{\Thr}_{X^n|S_nY^nZ^n}(X^n|S_nY^nZ^n)}
               {p_{X^n|S_nY^n}(X^n|S_nY^n)},
\nonumber\\
R+\eta&\geq&
\frac{1}{n}\log
\frac{q^{\Fou}_{X^n|S_nY^n}(X^n|S_nY^n)}{p_{X^n|Y^n}(X^n|Y^n)},
\nonumber\\
{\Dist} &\geq & 
\left.
\frac{1}{n}\log \exp \left\{ d(X^n,Z^n) \right\}
\right\}.
\label{eqn:azsadad2}
\eeqa
Hence, it suffices to show 
\beqno
& &{\rm P}_{\rm c}^{(n)}(\varphi_1^{(n)},\varphi_2^{(n)},\psi^{(n)};{\Dist})
\\
&\leq&
p_{S_nX^nY^n}\left({\cal A}_n
 \cap {\cal B}_n
 \cap {\cal C}_n \cap {\cal D}_n \cap {\cal E}_n\right)
 +4{\rm e}^{-n\eta}
\eeqno
to prove Lemma \ref{lm:Ohzzz}. By definition we have
\beqno 
&&{\rm P}_{\rm c}^{(n)}(\varphi^{(n)},\psi^{(n)};{\Dist})
=p_{S_nX^nY^nZ^n}\left({\cal E}_n\right).
\eeqno
Then we have the following.
\beqno
&&{\rm P}_{\rm c}^{(n)}(\varphi^{(n)},\psi^{(n)};{\Dist})
=p_{S_nX^nY^nZ^n}\left({\cal E}_n\right)
\\
&=&
p_{S_nX^nY^nZ^n}\left(
     {\cal A}_n
\cap {\cal B}_n
\cap {\cal C}_n
\cap {\cal D}_n
\cap {\cal E}_n
\right)
\\
& &
+p_{S_nX^nY^nZ^n}
\left(
\left[{\cal A}_n
 \cap {\cal B}_n
 \cap {\cal C}_n 
 \cap {\cal D}_n\right]^{\rm c}
 \cap {\cal E}_n 
\right)
\\
&\leq&
p_{S_nX^nY^nZ^n}
\left({\cal A}_n
 \cap {\cal B}_n
 \cap {\cal C}_n
 \cap {\cal D}_n
 \cap {\cal E}_n
\right)
\\
& &+p_{S_nX^nY^nZ^n}
\left(
{\cal A}_n^{\rm c}
\right)
+p_{S_nX^nY^nZ^n}
\left(
{\cal B}_n^{\rm c}
\right)
\\
& &
+p_{S_nX^nY^nZ^n}
\left(
{\cal C}_n^{\rm c}
\right)
+p_{S_nX^nY^nZ^n}\left(
{\cal D}_n^{\rm c} 
\right)
\\
&\MLeq{a}&
p_{S_nX^nY^nZ^n}\left(
     {\cal A}_n
\cap {\cal B}_n
\cap {\cal C}_n
\cap {\cal D}_n
\cap {\cal E}_n
\right)
+4{\rm e}^{-n\eta}. 
\eeqno
Step (a) follows from Lemma \ref{lm:zzxa}.
\hfill\IEEEQED
}
\begin{lm}\label{lm:Mchain} 
Suppose that for each $t=1,2,\cdots,n$, 
the joint distribution $p_{S_nX^tY^n}$ 
of the random vector $S_nX^tY^n$ is a marginal 
distribution of $p_{S_nX^nY^n}$. 
Then, for $t=1,2,\cdots,n$, we have the following 
Markov chain:
\beq
X_t \leftrightarrow S_nX^{t-1}Y_{t}^n \leftrightarrow Y^{t-1} 
\label{eqn:sssq} 
\eeq
or equivalently that $I(X_t;Y^{t-1}|S_nX^{t-1}Y_t^n)=0$. 
\end{lm}

Proof of this lemma is {\GorF} in Appendix \ref{sub:Apdb}.
\newcommand{\Apdb}{
\subsection{
Proof of Lemma \ref{lm:Mchain} 
}\label{sub:Apdb}

In this appendix we prove Lemma \ref{lm:Mchain}. 

{\it Proof of Lemma \ref{lm:Mchain}:} We have the following 
chain of inequalities:
\beqno
&  &I(X_t;Y^{t-1}|S_nX_{t-1}Y_t^n)
\\
&=&H(Y^{t-1}|S_nX^{t-1}Y_t^n)-H(Y^{t-1}|S_nX^{t}Y_t^n) 
\\
&\leq &H(Y^{t-1}|X^{t-1})-H(Y^{t-1}|S_n X^nY_t^n) 
\\
&\MEq{a}& H(Y^{t-1}|X^{t-1})-H(Y^{t-1}|X^n Y_t^n)
\\
&\MEq{b}& H(Y^{t-1}|X^{t-1})-H(Y^{t-1}|X^{t-1})=0.
\eeqno
Step (a) follows from that $S_n=\varphi^{(n)}(X^n)$ is 
a function of $X^n$. Step (b) follows 
from the memoryless property of 
the information source $\{(X_t,Y_t)\}_{t=1}^{\infty}$.
%
\hfill\IEEEQED
}
For $t=1,2, \cdots,n$, set $u_t \defeq (s,x^{t-1},y_{t+1}^n)$.
Let $U_t \defeq $ $(S_n,X^{t-1},$ $Y_{t+1}^n)$ be a random vector 
taking values in 
${\cal M}_n$
$\times$ ${\cal X}^{t-1}$ 
$\times$ ${\cal Y}_{t+1}^n$. From Lemmas \ref{lm:Ohzzz} 
and \ref{lm:Mchain}, we have 
the following. 
\begin{lm}\label{lm:Ohzzzaaa} For any $\eta>0$ 
and for any $(\varphi^{(n)}$, $\psi^{(n)})$ satisfying 
$$
\frac{1}{n} \log M_n \leq R,
$$
we have the following:
\beqa
\lefteqn{
\hspace*{-6mm}
{\rm P}_{\rm c}^{(n)}(\varphi^{(n)}, \psi^{(n)};{\Dist})
\leq p_{S_nX^nY^nZ^n}\biggl\{  
}
\nonumber\\
\eta & \geq& \frac{1}{n}\sum_{t=1}^n 
\log \frac{q^{\One}_{{X}_t}(X_t)}{p_{X_t}(X_t)},
\nonumber\\
\eta & \geq& \frac{1}{n}\sum_{t=1}^n \log 
\frac{q^{\Two}_{Y_t|U_tX_t}(Y_t|U_t,X_t)}
     {p_{Y_t|X_t}(Y_t|X_t)},
\nonumber\\
\eta & \geq & \frac{1}{n}\sum_{t=1}^n 
\log \frac{q^{\Thr}_{X_t|U_tY_tZ_t}(X_t|U_t,Y_t,Z_t)}
          {p_{X_t|U_tY_t}(X_t|U_t,Y_t)},
\nonumber\\
R+\eta & \geq & 
\frac{1}{n}\sum_{t=1}^n 
\log \frac{q^{\Fou}_{X_t|U_tY_t}(X_t|U_t,Y_t)}{p_{X_t|Y_t}(X_t|Y_t)},
\nonumber\\
{\Dist} &\geq & \left.
\frac{1}{n} \sum_{t=1}^n \log {\rm e}^{d(X_t,Z_t)}
\right\}
+4{\rm e}^{-n\eta},
\label{eqn:Zsdaa}
\eeqa
where 
for each $t=1,2,\cdots,n$, the following probability distribution
and stochastic matrices:
$$
q^{\One}_{X_t},
q^{\Two}_{Y_t|U_tX_t}, 
q^{\Thr}_{X_t|U_tY_tZ_t}, \mbox{ and }
q^{\Fou}_{X_t|U_tY_t}
$$
appearing in the first term in the right members of 
(\ref{eqn:Zsdaa}) have a property that 
we can choose their values arbitrary. 
\end{lm}

{\it Proof:} On the probability distributions appearing in 
the right members of (\ref{eqn:azsada}), we take the following 
choices. In (\ref{eqn:asppa}), we choose $q^{\One}_{X^n}$ so that 
\beq
q^{\One}_{X^n}(X^n)=\prod_{t=1}^n q^{\One}_{X_t}(X_t).
\label{eqn:ddsa}
\eeq
In (\ref{eqn:asppb}), we choose $q^{\Two}_{Y^n|S_nX^n}$ so that 
\beqa
& &q^{\Two}_{Y^n|S_nX^n}(Y^n|S_n,X^n)
\nonumber\\
&=&\prod_{t=1}^n q^{\Two}_{Y_t|S_nX^{t}Y_{t+1}^n}(Y_t|S_n,X^{t},Y_{t+1}^n)
\nonumber\\
&=&\prod_{t=1}^n q^{\Two}_{Y_t|X_{t}U_t}(Y_t|U_tX_{t}).
\label{eqn:ddsb}
\eeqa
In (\ref{eqn:asppc}), we choose $q^{\Thr}_{X^n|S_nY^nZ^n}$ so that 
\beqa
& &q^{\Thr}_{X^n|S_nY^nZ^n}(X^n|S_n,Y^n,Z^n)
\nonumber\\
&=&
\prod_{t=1}^n 
q^{\Thr}_{X_t|S_nX^{t-1}Y_{t}^nZ_t}(X_t|S_nX^{t-1},Y_t^n,Z_t)
\nonumber\\
&=&
\prod_{t=1}^n 
q^{\Thr}_{X_t|U_tY_tZ_t}(X_t|U_tY_tZ_t).
\label{eqn:ddsc}
\eeqa
In (\ref{eqn:asppc}), we note that 
\beqa
& &p_{X^n|S_nY^n}(X^n|S_n,Y^n)
\nonumber\\
&=&\prod_{t=1}^n p_{X_t|S_nX^{t-1}Y^n}(X_t|S_n,X^{t-1},Y^n)
\nonumber\\
&\MEq{a}&\prod_{t=1}^n p_{X_t|S_nX^{t-1}Y_t^n}(X_t|S_n,X^{t-1},Y_t^n)
\nonumber\\
&=&\prod_{t=1}^n p_{X_t|U_tY_t}(X_t|U_t,Y_t).
\label{eqn:ddscb}
\eeqa
Step (a) follows  from Lemma \ref{lm:Mchain}. In (\ref{eqn:asppd}), 
we choose $q^{\Fou}_{X^n|S_nY^n}$ so that 
\beqa
& &q^{\Fou}_{X^n|S_nY^n}(X^n|S_n,Y^n)
\nonumber\\
&=& 
\prod_{t=1}^n 
q^{\Fou}_{X_t|S_nX^{t-1}Y_{t}^n}(X_t|S_n,X^{t-1},Y_{t}^n)
\nonumber\\
&=& 
\prod_{t=1}^n 
q^{\Fou}_{X_t|U_tY_t}(X_t|U_t,Y_t).
\label{eqn:ddsd}
\eeqa
From Lemma \ref{lm:Ohzzz} and (\ref{eqn:ddsa})-(\ref{eqn:ddsd}), 
we have the bound (\ref{eqn:Zsdaa}) in Lemma \ref{lm:Ohzzzc}.
\hfill \IEEEQED

For each $t=1,2,\cdots,n$, let 
$
{\cal Q}({\cal U}_t $ 
$\times {\cal X} \times $
${\cal Y} \times {\cal Z})
$
be a set of all probability distributions $\ByShun{q_{U_tX_tY_tZ_t}}$ on
$$ 
{\cal U}_t \times {\cal X} \times 
   {\cal Y} \times {\cal Z}
={\cal M}_n \times {\cal X}^t \times {\cal Y}^{n-t+1} 
\times {\cal Z}
$$
\ByShun{such that the support of the marginal distribution 
$q_{X_tY_t}$ is included in that of $p_{XY}$.}
For $t=1,2,\cdots, n$, we simply write  
${\cal Q}_t$$=$${\cal Q}({\cal U}_t $ 
$\times {\cal X} \times {\cal Y} \times {\cal Z})$.
Similarly, for $t=1,2,\cdots, n$, we simply write 
$q_t=$$q_{U_tX_tY_tZ_t}$ $\in {\cal Q}_t$.
Set 
\beqno
{\cal Q}^n&\defeq& 
\prod_{t=1}^n {\cal Q}_t=\prod_{t=1}^n{\cal Q}(
{\cal U}_t \times {\cal X}\times{\cal Y}\times {\cal Z}),
\\
q^n & \defeq & \left\{ q_t \right\}_{t=1}^n \in {\cal Q}^n.
\eeqno
From Lemma \ref{lm:Ohzzzaaa}, we immediately obtain 
the following lemma. 
\begin{lm}\label{lm:Ohzzzc} For any $\eta>0$, 
for any $(\varphi^{(n)}$, $\psi^{(n)})$ satisfying 
$$
\frac{1}{n} \log M_n \leq R,
$$
and for any $q^n\in {\cal Q}^n$, 
we have the following:
\beqa
\lefteqn{
\hspace*{-6mm}
{\rm P}_{\rm c}^{(n)}(\varphi^{(n)}, \psi^{(n)};{\Dist})
\leq p_{S_nX^nY^nZ^n}\biggl\{  
}
\nonumber\\
\eta & \geq& \frac{1}{n}\sum_{t=1}^n 
\log \frac{q_{{X}_t}(X_t)}{p_{X_t}(X_t)},
\nonumber\\
\eta & \geq& \frac{1}{n}\sum_{t=1}^n \log 
\frac{q_{Y_t|U_tX_t}(Y_t|U_t,X_t)}
     {p_{Y_t|X_t}(Y_t|X_t)},
\nonumber\\
\eta & \geq & \frac{1}{n}\sum_{t=1}^n 
\log \frac{q_{X_t|U_tY_tZ_t}(X_t|U_t,Y_t,Z_t)}
          {p_{X_t|U_tY_t}(X_t|U_t,Y_t)},
\nonumber\\
R+\eta & \geq & 
\frac{1}{n}\sum_{t=1}^n 
\log \frac{q_{X_t|U_tY_t}(X_t|U_t,Y_t)}{p_{X_t|Y_t}(X_t|Y_t)},
\nonumber\\
{\Dist} &\geq & \left.
\frac{1}{n} \sum_{t=1}^n \log {\rm e}^{d(X_t,Z_t)}
\right\}
+4{\rm e}^{-n\eta},
\label{eqn:aazsadz}
\eeqa
where for each $t=1,2,\cdots,n$, the following probability distribution
and stochastic matrices:
$$
q_{X_t},
q_{Y_t|U_tX_t}, 
q_{X_t|U_tY_tZ_t}, \mbox{ and }
q_{X_t|U_tY_t}
$$
appearing in the first term in the right members of (\ref{eqn:aazsadz})
are chosen so that they are induced by the joint distribution 
$q_t=q_{U_tX_tY_tZ_t} \in {\cal Q}_t$.
\end{lm}

To evaluate an upper bound of 
(\ref{eqn:aazsadz}) in Lemma \ref{lm:Ohzzzc}. 
We use the following lemma, which is well known as the Cram\ByShun{\'e}r's bound in 
the large deviation principle.
\begin{lm}
\label{lm:Ohzzzb}
For any real valued random variable 
$Z$ and any $\theta>0$, we have
$$
\Pr\{Z \geq a \}\leq 
\exp
\left[
-\left(
\theta a -\log {\rm E}[\exp(\theta Z)]
\right) 
\right].
$$
\end{lm}

Here we define a quantity which serves as an exponential
upper bound of ${\rm P}_{\rm c}^{(n)}(\varphi^{(n)},$ 
$\psi^{(n)})$. 
Let ${\cal P}^{(n)}(p_{XY})$ be a 
set of all probability distributions 
${p}_{S_nX^nY^nZ^n}$ on 
${\cal M}_n$
$\times {\cal X}^n$
$\times {\cal Y}^n$
$\times {\cal Z}^n$
having the form:
\beqno
& &{p}_{S_nX^nY^nZ^n}(s,x^n,y^n,z^n)
\\
&=&{p}_{S_n|X^n}(s|x^n)
\left\{\prod_{t=1}^n 
p_{X_tY_t}(x_t,y_t)\right\}
p_{Z^n|Y^nS_n}(z^n|y^n,s).
\eeqno
For simplicity of notation we use the notation $p^{(n)}$ 
for $p_{S_nX^nY^nZ^n}$ $\in {\cal P}^{(n)}$
$(p_{XY})$. We assume that 
$
p_{U_tX_tY_tZ_t}=p_{S_nX^{t}Y_t^n Z_t}
$
is a marginal distribution of $p^{(n)}$.
For $t=1,2,\cdots, n$, we simply write $p_t=$ $p_{U_tX_tY_tZ_t}$. 
For $p^{(n)}$ $\in {\cal P}^{(n)}(p_{XY})$ 
and $q^n$ $\in {\cal Q}^n$, we define 
\beqno
& &
\Omega^{(\alpha,\mu, \theta)}_{\pNqN}
{\ARgRv}
\\
&\defeq & 
-\log {\rm E}_{p^{(n)}}\left[
\prod_{t=1}^n\right.
\exp \left\{ -\theta 
\omega^{(\ByShun{\alpha,\mu})}_{p_{t}||q_{t}}
(X_t,Y_t,Z_t|U_t) \right\}
\HugebrB
\\
&=&
-\log
{\rm E}_{p^{(n)}}
\left[
\left\{
\prod_{t=1}^n
\frac 
{p_{X_t}^{\bar{\alpha}\theta}(X_t)}
{q_{X_t}^{\bar{\alpha}\theta}(X_t)}
\frac 
{p^{\bar{\alpha}\theta}_{Y_t|X_t}(Y_t|X_t)}
{q^{\bar{\alpha}\theta}_{Y_t|X_tU_t}(Y_t|X_t,U_t)}
\right.\right\}
\\
& &\times \left\{
\prod_{t=1}^n
\frac 
{p^{\bar{\alpha}\theta}_{X_t|U_tY_t}(X_t|U_t,Y_t)}
{q^{\bar{\alpha}\theta}_{X_t|U_tY_tZ_t}(X_t|U_t,Y_t,Z_t)}
\right\}
\\
& &\times \left\{
\left.
\prod_{t=1}^n
\frac
{p^{4\alpha\theta}_{X_t|Y_t}(X_t|Y_t)}
{q^{4\alpha\theta}_{X_t|Y_tU_t}(X_t|U_t,Y_t)}
{\rm e}^{4\alpha \mu \theta d(X_t,Z_t)}\right\}
\right],
\eeqno
where for each $t=1,2,\cdots,n$, the following probability 
distribution and stochastic matrices:
$$
q_{X_t},
q_{X_t|U_tY_t}, 
q_{X_t|U_tY_tZ_t}, 
q_{Y_t|X_tU_t}
$$ 
appearing in the definition of 
$
\Omega^{(\alpha,\mu,\theta)}_{\pNqN}
{\ARgRv}
$
are chosen so that they are induced by the joint distribution 
$q_t=q_{U_tX_tY_tZ_t}\in {\cal Q}_t$. 

Here we give a remark on an essential difference 
between $p^{(n)}$ $\in {\cal P}^{(n)}(p_{XY})$ 
and $q^n$ $\in {\cal Q}^n$. 
For the former the $n$ probability distributions 
$p_t,$ $t=1,2,\cdots, n,$ are consistent with $p^{(n)}$, 
since they are marginal distributions of $p^{(n)}$. 
On the other hand, for the latter, $q^{n}$ 
is just {\it a sequence} of $n$ 
probability distributions. Hence, we may not have the 
consistency between the $n$ elements $q_t$, $t=1,2,\cdots,n,$ 
of $q^n$. 

By Lemmas \ref{lm:Ohzzzc} and \ref{lm:Ohzzzb}, we have the 
following proposition. 
\begin{pro}
\label{pro:Ohzzp}
For any 
$\alpha,$ 
$\mu,$ 
$\theta >0$, any $q^n \in {\cal Q}^n$, and 
any $(\varphi^{(n)},$ $\psi^{(n)})$ satisfying 
$$
\frac{1}{n} \log {M_n}\leq R,  
$$
we have 
\beqno
& &{\rm P}_{\rm c}^{(n)}(\varphi^{(n)},\psi^{(n)};{\Dist})
\\
&\leq &5\exp
\biggl\{
-n\left[1+(3+\alpha)\theta\right]^{-1}
\\
& & \times 
\left. \left[
\frac{1}{n}
\Omega^{(\alpha, \mu,\theta)}_{\pNqN}{\ARgRv}
-4 \alpha \theta(\bar{\mu} R+\mu{\Dist})\right]
\right\}.
\eeqno
\end{pro}

{\it Proof:} By Lemma \ref{lm:Ohzzzc}, for $\alpha,$ $\mu,\theta$ $\geq 0$, 
we have the following 
chain of inequalities: 


\beqa
& &
\lefteqn{\hspace*{-6mm}
{\rm P}_{\rm c}^{(n)}(\varphi^{(n)}, \psi^{(n)};{\Dist})
\leq p_{S_nX^nY^nZ^n}\biggl\{  
}
\nonumber\\
& &
\ba{rcl}
\bar{\alpha}\eta & \geq& \ds \frac{\bar{\alpha}}{n}\sum_{t=1}^n 
\log \frac{q_{{X}_t}(X_t)}{p_{X_t}(X_t)},
\vspace*{2mm}\\
\bar{\alpha}\eta & \geq& \ds \frac{\bar{\alpha}}{n}\sum_{t=1}^n \log 
\frac{q_{Y_t|U_tX_t}(Y_t|U_t,X_t)}
     {p_{Y_t|X_t}(Y_t|X_t)},
\vspace*{2mm}\\
\bar{\alpha}\eta & \geq &\ds  \frac{\bar{\alpha}}{n}\sum_{t=1}^n 
\log \frac{q_{X_t|U_tY_tZ_t}(X_t|U_t,Y_t,Z_t)}
          {p_{X_t|U_tY_t}(X_t|U_t,Y_t)},
\vspace*{2mm}\\
4\alpha\bar{\mu}(R+\eta) & \geq &\ds 
\frac{4\alpha\bar{\mu}}{n}\sum_{t=1}^n 
\log \frac{q_{X_t|U_tY_t}(X_t|U_t,Y_t)}{p_{X_t|Y_t}(X_t|Y_t)},
\vspace*{2mm}\\
4\alpha\mu{\Dist} &\geq &\ds \left.
\frac{4\alpha\mu}{n} \sum_{t=1}^n \log {\rm e}^{d(X_t,Z_t)}
\right\}+4{\rm e}^{-n\eta}
\ea
\nonumber\\
&\leq & p_{S_nX^nY^nZ^n}\hugel
4\alpha(\bar{\mu}R+\mu{\Dist})+(3+\alpha-4\alpha\mu)\ByShun{\eta} 
\nonumber\\
& &
\left. \geq 
\frac{1}{n}
\sum_{t=1}^n
\omega_{{\ptANDqt}}^{(\alpha,\mu)}
(X_t,Y_t,Z_t|U_t)
\right\}
+4{\rm e}^{-n\eta} 
\nonumber\\
&=&p_{S_nX^nY^nZ^n}\left\{
-\frac{1}{n}
\sum_{t=1}^n
\theta\omega_{{\ptANDqt}}^{(\alpha,\mu)}(X_t,Y_t,Z_t|U_t)
\right.
\nonumber\\
& &
\geq - \theta 
\left[4\alpha(\bar{\mu}R+\mu{\Dist})+(3+\alpha-4\alpha\mu)\eta\right] \biggr\}
+4{\rm e}^{-n\eta} 
\nonumber\\
&\MLeq{a} &
\exp \biggl[n\biggl\{4\alpha\theta(\bar{\mu} R+\mu{\Dist})
+(3+\alpha-4\alpha\mu)\theta\eta 
\nonumber\\
&&\left.\left.-\frac{1}{n}
\Omega^{(\alpha,\mu,\theta)}_{\pNqN}
{\ARgRv}
\right\}\right]
+4{\rm e}^{-n\eta}.\label{eqn:aaabv}
\eeqa
Step (a) follows from Lemma \ref{lm:Ohzzzb}. 
We choose $\eta$ so that 
\beqa
-\eta&=& 4\alpha \theta(\bar{\mu} R+\mu{\Dist})
+\theta(3+\alpha-4\alpha\mu)\eta
\nonumber\\ 
&  &-\frac{1}{n}
\Omega^{(\alpha,\mu,\theta)}
_{\pNqN}{\ARgRv}.
\label{eqn:aaappp}
\eeqa
Solving (\ref{eqn:aaappp}) with respect to $\eta$, we have 
\beqno
\eta=
\frac{\frac{1}{n}
\Omega^{(\alpha,\mu,\theta)}_{\pNqN}{\ARgRv}
-4\alpha\theta(\bar{\mu}R+\mu{\Dist})}
{1+(3+\alpha-4\alpha\mu)\theta}.
\eeqno
For this choice of $\eta$ and (\ref{eqn:aaabv}), we have
\beqno
& &{\rm P}_{\rm c}^{(n)}(\varphi^{(n)},\psi^{(n)};{\Dist})
\leq 5{\rm e}^{-n\eta}
\\
&=&
5\exp
\biggl\{
-n\left[1+(3+\alpha-4\alpha\mu)\theta \right]^{-1}
\\
& & \times 
\left. \left[
\frac{1}{n}\Omega^{(\alpha,\mu,\theta)}_{\pNqN}
{\ARgRv}-4\alpha\theta(\bar{\mu} R+\mu{\Dist})\right]
\right\},
\eeqno
completing the proof. 
\hfill \IEEEQED

Set 
\beqno
& &\underline{\Omega}^{(\alpha, \mu, \theta)}(p_{XY})
\\
&\defeq & 
\inf_{n\geq 1}
\min_{
p^{(n)}\in {\cal P}^{(n)}(p_{XY}) 
}
\max_{\scs q^n \in {\cal Q}^n}
\frac{1}{n}
\Omega^{(\alpha, \mu,\theta)}_{\pNqN}
{\ARgRv}.
\eeqno
By Proposition \ref{pro:Ohzzp} we have the following corollary.  
\begin{co} \label{co:corOne}
For any $(\alpha,\mu)\in [0,1]^2$, for any $\theta >0$, and for any 
$(\varphi^{(n)},$ $\psi^{(n)})$ 
satisfying 
$$
\frac{1}{n}\log ||\varphi^{(n)}||\leq R, 
$$
we have 
\beqno
& &{\rm P}_{\rm c}^{(n)}(\varphi^{(n)},\psi^{(n)};{\Dist})
\\
&\leq &5\exp
\left\{-n\left[
\frac{\underline{\Omega}^{(\alpha,\mu,\theta)}(p_{XY})
-4\alpha\theta(\bar{\mu}R+ \mu{\Dist})}
{1+(3+\alpha-4\alpha\mu)\theta}
\right]\right\}.
\eeqno
\end{co}

We shall call 
$\underline{\Omega}^{(\alpha,\mu,\theta)}(p_{XY})$ 
the communication potential. The above corollary implies that 
the analysis of 
$\underline{\Omega}^{(\alpha,\mu,\theta)}($
$p_{XY})$ leads to an establishment of a strong converse
theorem for Wyner-Ziv source coding problem. In the
following argument we drive an explicit lower bound of
$\underline{\Omega}^{(\alpha,\mu,\theta)}(p_{XY})$. 
We use a \ByShun{new} techique we call {\it the recursive method}. 
The recursive method is a powerfull tool to drive a single 
letterized exponent function for rates below the 
rate distortion function. This method is also applicable 
to prove the exponential strong converse theorems for other 
network information theory problems 
\cite{OhIsit15AKWStConv}, 
\cite{OhIsit15DBCStConv}, 
\cite{OhIsit15DBCFBStConv}.

\newcommand{\oMegga}{
We define
\beqno
& &
\Omega^{(\alpha,\mu, \theta)}_{\pNqN}
{\ARgRv}
\\
&\defeq & 
-\log {\rm E}\left[
\prod_{t=1}^n\right.
\exp \left\{
-\theta
\omega^{(\ByShun{\alpha,\mu})}_{p_{t}||q_{t}}
(X_tY_tZ_t|U_t) \right\}
\HugebrB
\\
&=&
-\log
{\rm E}
\left[
\left\{
\prod_{t=1}^n
\frac 
{p_{X_t}^{\bar{\alpha}\theta}(X_t)}
{q_{X_t}^{\bar{\alpha}\theta}(X_t)}
\frac 
{p^{\bar{\alpha}\theta}_{Y_t|X_t}(Y_t|X_t)}
{q^{\bar{\alpha} \theta}_{Y_t|X_tU_t}(Y_t|X_t,U_t)}
\right.\right\}
\\
& &\times \left\{
\prod_{t=1}^n
\frac 
{p^{\bar{\alpha} \theta}_{X_t|U_tY_t}(X_t|U_t,Y_t)}
{q^{\bar{\alpha} \theta}_{X_t|U_tY_tZ_t}(X_t|U_t,Y_t,Z_t)}
\right\}
\\
& &\times \left\{
\left.
\prod_{t=1}^n
\frac
{p^{4\alpha \theta}_{X_t|Y_t}(X_t|Y_t)}
{q^{4\alpha \theta}_{X_t|Y_tU_t}(X_t|U_t,Y_t)}
{\rm e}^{\mu d(X_t,Z_t)}\right\}
\right].
\eeqno
}

For each $t=1,2,\cdots,n$, set
\beqno
&&
f_{{\ptANDqt}}^{(\alpha,\mu,\theta)}
(x_t,y_t,z_t|u_t)
\nonumber\\
&\defeq & 
\exp 
\left\{ 
-\theta \omega^{(\ByShun{\alpha,\mu})}_{{\ptANDqt}}
(x_t,y_t,z_t|u_t)
\right\}
\\
&=&
\frac{p_{X_t}^{\bar{\alpha}\theta}(x_t)}
{q_{X_t}^{\bar{\alpha}\theta}(x_t)}
\frac{p^{\bar{\alpha} \theta}_{Y_t|X_t}(y_t|x_t)}
{q^{\bar{\alpha} \theta}_{Y_t|X_tU_t}(y_t|x_t,u_t)}
\frac 
{p^{\bar{\alpha} \theta}_{X_t|U_tY_t}(x_t|u_t,y_t)}
{q^{\bar{\alpha} \theta}_{X_t|U_tY_tZ_t}(x_t|u_t,y_t,z_t)}
\\
& &\times 
\frac
{p^{4 \alpha\theta}_{X_t|Y_t}(x_t|y_t)}
{q^{4 \alpha\theta}_{X_t|Y_tU_t}(x_t|u_t,y_t)}
{\rm e}^{4\alpha \mu\theta d(x_t,z_t)}.
\eeqno
By definition we have
\beqa
&&
\exp\left\{-
\Omega^{(\alpha,\mu, \theta)}_{\pNqN}
{\ARgRv}
\right\}
\nonumber\\
&=&\sum_{s,y^n}p_{S_nY^n}(s,y^n)
\sum_{x^n,z^n}p_{X^nZ^n|S_nY^n}(x^n,z^n|s,y^n)
\nonumber\\
& &\qquad \qquad \times \prod_{t=1}^n
f_{{\ptANDqt}}^{(\alpha,\mu,\theta)}
(x_t,y_t,z_t|u_t).
\label{eqn:aZssf}
\eeqa
For each $t=1,2, \cdots, n$, we define the conditional probability 
distribution
\beqno
& & 
{p}_{X^tZ^t|S_nY^n}^{(\alpha,\mu,\theta;q^t)}
\\
&\defeq &
\left\{
p_{X^tZ^t|S_nY^n}^{(\alpha,\mu,\theta;q^t)}
(x^t,z^t|s,y^n)
\right\}_{
(x^t,z^t,s,y^n)
   \in {\cal X}^t \times {\cal Z}^t
\times {\cal M}_n \times {\cal Y}^n}
\eeqno
by
\beqno
& &
p_{X^tZ^t|S_nY^n}^{(\alpha,\mu,\theta;q^t)}
(x^t,z^t|s,y^n)
\\ 
&\defeq& 
C_t^{-1}(s,y^n)
p_{X^tZ^t|S_nY^n}(x^t,z^t|s,y^n)
\\
& &\quad \times 
\prod_{i=1}^t
f_{\ByShun{q_i||p_i}}^{(\alpha,\mu,\theta)}
(x_i,y_i,z_i|u_i)
\eeqno 
where
\beqa
C_t(s,y^n)
& \defeq & \sum_{x^t,z^t}p_{X^tZ^t|S_nY^n}(x^t,z^t|s,y^n)
\nonumber\\
& &\quad \times 
\prod_{i=1}^t
f_{\ByShun{q_i||p_i}}^{(\alpha,\mu,\theta)}
(x_i,y_i,z_i|u_i)
\label{eqn:Szzz}
\eeqa
are constants for normalization. For $t=1,2,\cdots,n$, define
\beq
\Phi_{t,q^t}^{(\alpha,\mu,\theta)}(s,y^n)
\defeq C_t(s,y^n)C_{t-1}^{-1}(s,y^n),
\label{eqn:defa}
\eeq
where we define $C_{0}(s,y^n)=1$ for 
$(s,y^n)\in {\cal M}_n $ $\times {\cal Y}^n.$
Then we have the following lemma.
\begin{lm}\label{lm:aaa}
For each $t=1,2,\cdots,n$, and for any 
$(s,$ $y^n$ $x^t, z^t)\in {\cal M}_n$
$\times {\cal Y}^n$
$\times {\cal X}^t$
$\times {\cal Z}^t$,
we have
\beqa
& &{p}_{X^tZ^t|S_nY^n}^{(\alpha,\mu,\theta;q^t)}
(x^t,z^t|s,y^n)
=(\Phi_{t,q^t}^{(\alpha,\mu,\theta)}(s,y^n))^{-1}
\nonumber\\
&& \quad \times
 p_{X^{t-1}Y^{t-1}|S_nY^n}^{(\alpha,\mu,\theta;q^{t-1})}
(x^{t-1},z^{t-1}|s,y^n)
\nonumber\\
& &\quad \times p_{X_tZ_t|S_nX^{t-1}Y^n}(x_t,z_t| s,x^{t-1},z^{t-1},y^n)
\nonumber\\
& &\quad \times 
f_{{\ptANDqt}}^{(\alpha,\mu,\theta)}(x_t,y_t,z_t|u_t).
\label{eqn:satt}
\\
& &
\Phi_{t,q^t}^{(\alpha,\mu,\theta)}(s,y^n)
\nonumber\\
&=& \sum_{x^t,z^t} 
p_{X^{t-1}Z^{t-1}|S_nY^n}^{(\alpha,\mu,\theta;q^{t-1})}
(x^{t-1},z^{t-1}|s,y^n)
\nonumber\\
& &\quad \times p_{X_tZ_t|S_nX^{t-1}Y^n}(x_t,z_t| s,x^{t-1},z^{t-1},y^n)
\nonumber\\
& &\quad\times 
f_{\ByShun{q_{t}||p_{t}}}^{(\alpha,\mu,\theta)}
(x_t,y_t,z_t|u_t).
\label{eqn:sattx}
\eeqa
Furthermore, we have  
\beqa
&& 
\exp\left\{-
\Omega^{(\alpha,\mu, \theta)}_{\pNqN}
{\ARgRv}
\right\}
\nonumber\\
&=&
\sum_{s,y^n}p_{S_nY^n}(s,y^n) 
\prod_{t=1}^n \Phi_{t,q^t}^{(\alpha,\mu,\theta)}(s,y^n).
\label{eqn:sattxb}
\eeqa
\end{lm}

The equality (\ref{eqn:sattxb}) in Lemma \ref{lm:aaa} is 
obvious from (\ref{eqn:aZssf}), (\ref{eqn:Szzz}), and (\ref{eqn:defa}).
Proofs of (\ref{eqn:satt}) and (\ref{eqn:sattx}) in this lemma are 
given in Appendix \ref{sub:sdfa}.
\newcommand{\Apdc}{
\subsection{Proof of Lemma \ref{lm:aaa}}\label{sub:sdfa}
In this appendix we prove 
(\ref{eqn:satt}) and (\ref{eqn:sattx}) in
Lemma \ref{lm:aaa}.  

{\it Proofs of (\ref{eqn:satt}) and (\ref{eqn:sattx}) in Lemma \ref{lm:aaa}: }By 
the definition of ${p}_{X^tZ^t|S_nY^n,q^t}^{(\alpha,\mu,\theta)}$ $(x^t,z^t|s,y^n)$, 
for $t=1,2,\cdots,n$, we have 
\beqa
& &p_{X^tZ^t|S_nY^n}^{(\alpha,\mu,\theta;q^{t})}(x^t,z^t|s,y^n)
\nonumber\\
&=&C_t^{-1}(s,y^n)
p_{X^tZ^t|S_nY^n}(x^t,z^t|s,y^n) 
\nonumber\\
& &\times
\prod_{i=1}^t
f_{p_i||q_i}^{(\alpha,\mu,\theta)}
(x_i,y_i,z_i|u_i).
\label{eqn:azaq}
\eeqa 
Then we have the following chain of equalities:
\beqa 
& &
p_{X^tZ^t|S_nY^n}^{(\alpha,\mu,\theta;q^{t})}
(x^t,z^{t}|s,y^n)
\nonumber\\
&\MEq{a}&
C_t^{-1}(s,y^n)p_{X^tZ^t|S_nY^n}(x^t,z^t|s,y^n) 
\nonumber\\
& &\times \prod_{i=1}^t f_{p_i||q_i}^{(\alpha,\mu,\theta)}
(x_i,y_i,z_i|u_i)
\nonumber\\
&=&C_t^{-1}(s,y^n)p_{X^{t-1}Z^{t-1}|S_nY^n}(x^{t-1},z^{t-1}|s,y^n)
\nonumber\\
& &\times \prod_{i=1}^{t-1}
f_{p_i||q_i}^{(\alpha,\mu,\theta)}
(x_i,y_i,z_i|u_i)
\nonumber\\
& &
\times p_{ X_t|Z_t|X^{t-1}Z^{t-1}SY^n}(x_t,z_t|x^{t-1},z^{t-1},s,y^n) 
\nonumber\\
& & 
\times 
f_{{\ptANDqt}}^{(\alpha,\mu,\theta)}
(x_t,y_t|u_t)
\nonumber\\
&\MEq{b}&
\frac{C_{t-1}(s,y^n)}{C_t(s,y^n)}
p_{X^{t-1}Z^{t-1}|S_nY^n}^{(\alpha,\mu,\theta;q^{t-1})}
(x^{t-1},z^{t-1}|s,y^n)
\nonumber\\
& &
\times p_{X_t|Z_t|X^{t-1}Z^{t-1}S_nY^n}(x_t,z_t|x^{t-1},z^{t-1},s,y^n) 
\nonumber\\
& & 
\times 
f_{{\ptANDqt}}^{(\alpha,\mu,\theta)}
(x_t,y_t,z_t|u_t)
\nonumber\\
&=&(\Phi_t^{(\alpha,\mu,\theta)}(s,y^n))^{-1}
p_{X^{t-1}Z^{t-1}|S_nY^n}^{(\alpha,\mu,\theta;q^{t-1})}
(x_t,y_t,z_t|u_t)
\nonumber\\
& &
\times p_{X_t|Z_t|X^{t-1}Z^{t-1}S_nY^n}(x_t,z_t|x^{t-1},z^{t-1},s,y^n) 
\nonumber\\
& &
\times f_{{\ptANDqt}}^{(\alpha,\mu,\theta)}
(x_t,y_t, z_t|u_t).
\label{eqn:daaaq}
\eeqa
Steps (a) and (b) follow from (\ref{eqn:azaq}). 
From (\ref{eqn:daaaq}), we have 
\beqa 
& &\Phi_{t,q^t}^{(\alpha,\mu,\theta)}(s,y^n)
p_{X^tZ^t|S_nY^n}^{(\alpha,\mu,\theta)}
(x^t,z^t|s,y^n)
\label{eqn:daxx}
\\
&=&p_{X^{t-1}Z^{t-1}|S_nY^n}^{(\alpha,\mu,\theta;q^{t-1})}
(x^{t-1},z^{t-1}|s,y^n)
\nonumber\\
& &\times
p_{X_tZ_t|X^{t-1}Z^{t-1}S_nY^n}(x_t,z_t|x^{t-1},z^{t-1},s,y^n)
\nonumber\\
& &\times
f_{{\ptANDqt}}^{(\alpha,\mu,\theta)}
(x_t,y_t,z_t|u_t).
\label{eqn:daaxx}
\eeqa
Taking summations of (\ref{eqn:daxx}) and 
(\ref{eqn:daaxx}) with respect to $x^t,z^t$, 
we obtain 
\beqno 
& &\Phi_{t,q^t}^{(\alpha,\mu,\theta)}(s,y^n)
\\
&=&\sum_{x^t,z^t}
p_{X^{t-1}Z^{t-1}|S_nY^n }^{(\alpha,\mu,\theta;q^{t-1})}
(x^{t-1},z^{t-1}|s,y^n)
\nonumber\\
& &\times
p_{X_tZ_t|X^{t-1}Z^{t-1}S_nY^n}(x_t,z_t|x^{t-1},z^{t-1},s,y^n)
\nonumber\\
& &\times
f_{{\ptANDqt}}^{(\alpha,\mu,\theta)}
  (x_t,y_t,z_t|u_t),
\eeqno
completing the proof.
\hfill \IEEEQED
}
Next we define a probability distribution of the random pair 
$(S_n,Y^n)$ taking values in ${\cal M}_n$ $\times {\cal Y}^n$ by 
\beqa
& &p_{S_nY^n}^{(\alpha,\mu,\theta;q^t)}(s,y^n)
\nonumber\\
&=&\tilde{C}_t^{-1} 
p_{S_nY^n}(s,y^n)
\prod_{i=1}^t 
\Phi_{i,q^i}^{(\alpha,\mu,\theta)}(s,y^n),
\label{eqn:defzz}
\eeqa
where $\tilde{C}_t$ is a constant for normalization given by 
$$
\tilde{C}_t=\sum_{s,y^n}p_{S_nY^n}(s,y^n)
\prod_{i=1}^t 
\Phi_{i,q^i}^{(\alpha,\mu,\theta)}(s,y^n).
$$
For $t=1,2,\cdots,n$, define 
\beq
\Lambda_{t,q^t}^{(\alpha,\mu,\theta)}
\defeq \tilde{C_t}\tilde{C}_{t-1}^{-1},
\label{eqn:passdf}
\eeq
where we define $\tilde{C}_{0}=1$. Then we have the following.
\begin{lm}\label{lm:keylm}
\beqa
& &
\exp\left\{-
\Omega^{(\alpha,\mu, \theta)}_{\pNqN}
{\ARgRv}\right\}
=\prod_{t=1}^n 
\Lambda_{t,q^t}^{(\alpha,\mu,\theta)},
\quad\label{eqn:uzapa}
\\
& &\Lambda_{t,q^t}^{(\alpha,\mu,\theta)}
=\sum_{s,y^n}
p_{S_nY^n}^{(\alpha,\mu,\theta;q^{t-1})}(s,y^n)
\Phi_{t,q^t}^{(\alpha,\mu,\theta)}(s,y^n)
\nonumber\\
& &=\sum_{s,y^n}
p_{S_nY^n}^{(\alpha,\mu,\theta;q^{t-1})}(s,y^n)
\nonumber\\
& &\times
\sum_{x^t,z^t} 
p_{X^{t-1}Z^{t-1}|S_nY^n}^{(\alpha,\mu,\theta;q^{t-1})}
(x^{t-1},z^{t-1}|s,y^n)
\nonumber\\
& &\qquad \times 
p_{X_tZ_t|X^{t-1}Z^{t-1}S_nY^n}(x_t,z_t|x^{t-1},z^{t-1},s,y^n) 
\nonumber\\
& &\qquad \times 
f_{{\ptANDqt}}^{(\alpha,\mu,\theta)}(x_t,y_t,z_t|u_t). 
\label{eqn:uzapazz}
\eeqa
\end{lm}

{\it Proof}: By the equality (\ref{eqn:sattxb}) in Lemma \ref{lm:aaa}, we have  
\beqa
& &\exp\left\{-
   \Omega^{(\alpha,\mu, \theta)}_{\pNqN}
   {\ARgRv}\right\}
\nonumber\\
&=&\tilde{C}_n
=
\prod_{t=1}^n\tilde{C}_t\tilde{C}_{t-1}^{-1}
\MEq{a}\prod_{t=1}^n
\Lambda_{t,q^t}^{(\alpha,\mu,\theta)}. 
\label{eqn:zza}
\eeqa
Step (a) follows from the definition (\ref{eqn:passdf}) of 
$\Lambda_{t,q^t}^{(\alpha,\mu,\nu,\theta)}.$ 
We next prove  (\ref{eqn:uzapazz}) 
in Lemma \ref{lm:keylm}. Multiplying 
$\Lambda_{t,q^t}^{(\alpha,\mu,\theta)}
=$ $\tilde{C}_{t}/\tilde{C}_{t-1}$ to both sides of (\ref{eqn:defzz}),
we have 
\beqa
& &\Lambda_{t,q^t}^{(\alpha,\mu,\theta)}
 p_{S_nY^n}^{(\alpha,\mu,\theta;q^t)}(s,y^n)
\label{eqn:aadff}
\\
&=&\tilde{C}_{t-1}^{-1}
p_{S_nY^n}(s,y^n)
\prod_{i=1}^t 
\Phi_{i,\ByShun{q^i}}^{(\alpha,\mu,\theta)}(s,y^n),
\nonumber\\
&=&
p_{S_nY^n}^{(\alpha,\mu,\theta;q^{t-1})}(s,y^n)
\Phi_{t,q^t}^{(\alpha,\mu,\theta)}(s,y^n).
\label{eqn:aadffsss}
\eeqa
Taking summations of (\ref{eqn:aadff}) and (\ref{eqn:aadffsss}) 
with respect to $(s,y^n)$, we have (\ref{eqn:uzapazz}) 
in Lemma \ref{lm:keylm}. 
\hfill \IEEEQED

The following proposition is a mathematical core to prove 
our main result. 
\begin{pro}\label{pro:mainpro} For $\theta\in (0,1/\bar{\alpha})$, 
we choose the parameter $\lambda$ such that
\beq
\lambda=\frac{\theta}{1-\bar{\alpha}\theta} 
\Leftrightarrow \theta=\frac{\lambda}{1+\bar{\alpha} \lambda}. 
\label{eqn:Zqdd}
\eeq
Then for any $\alpha,\mu >0$ and 
for any $\theta\in (0,1/\bar{\alpha})$, we have 
\beq
\underline{\Omega}^{(\alpha, \mu,\theta)}(p_{XY})
\geq \frac{1}{1+\lambda\bar{\alpha}}
\Omega^{(\alpha, \mu,\lambda)}(p_{XY}).
\label{eqn:zzar}
\eeq
\end{pro}

{\it Proof:} Set
\beqno
& &\hat{\cal Q}_n
\defeq 
\{q=q_{UXYZ}: 
\pa {\cal U} \pa \leq 
\pa {\cal M}_n \pa \pa {\cal X}^{n-1}\pa \pa {\cal Y}^{n-1}\pa\},
\\
& &\hat{\Omega}_n^{(\alpha, \mu,\lambda)}(p_{XY})
\defeq
\min_{\scs 
     \atop{\scs 
     q\in \hat{\cal Q}_n
     }
}
\Omega^{(\alpha,\mu,\lambda)}(q|p_{XY}).
\eeqno
Set 
\beqa
& &p_{S_nX^tY_t^nZ_t}^{(\alpha,\mu,\theta;q^{t-1})}
(s,x^t,y_t^n,z_t)
\nonumber\\
&=&p_{U_tX_tY_tZ_t}^{(\alpha,\mu,\theta;q^{t-1})}
(u_t,x_t,y_t,z_t)
\nonumber\\
&\defeq&\sum_{y^{t-1},z^{t-1}}
p_{S_nY^n}^{(\alpha,\mu,\theta;q^{t-1})}(s,y^n)
\nonumber\\
& &\times
p_{X^{t-1}Z^{t-1}|S_nY^n}^{(\alpha,\mu,\theta;q^{t-1})}
(x^{t-1},z^{t-1}|s,y^n)
\nonumber\\
& &\times 
p_{X_tZ_t|X^{t-1}Z^{t-1}S_nY^n}(x_t,z_t|x^{t-1},z^{t-1},s,y^n). 
\eeqa
Then by Lemma $\ref{lm:keylm}$, we have
\beqa
\Lambda_{t,q^t}^{(\alpha,\mu,\theta)}
&=& \sum_{u_t,x_t,y_t,z_t }
p_{U_tX_tY_tZ_t}^{(\alpha,\mu,\theta;q^{t-1})}
(u_t,x_t,y_t,z_t) 
\nonumber\\
& &\qquad\qquad \times
f_{{\ptANDqt}}^{(\alpha,\mu,\theta)}(x_t,y_t,z_t|u_t). 
\nonumber
\eeqa
For each $t=1,2,\cdots,n$, we \ByShun{recursively} choose 
$q_t=q_{U_tX_tY_tZ_t}$ so that 
$$
q_{U_tX_tY_tZ_t}(u_t,x_t,y_t,z_t)
=p_{U_tX_tY_tZ_t}^{(\alpha,\mu,\theta;q^{t-1})}
(u_t,x_t,y_t,z_t)
$$
and choose 
$q_{X_t}$, 
$q_{Y_t|X_tU_t}$, 
$q_{X_t|U_tY_tZ_t}$, and $q_{X_t|Y_tU_t}$ 
appearing in
\beqno
&&
f_{{\ptANDqt}}^{(\alpha,\mu,\theta)}
(x_t,y_t,z_t|u_t)
=
\frac{p_{X_t}^{\bar{\alpha}\theta}(x_t)}
{q_{X_t}^{\alpha\theta}(x_t)}
\frac{p^{\bar{\alpha} \theta}_{Y_t|X_t}(y_t|x_t)}
{q^{\bar{\alpha} \theta}_{Y_t|X_tU_t}(y_t|x_t,u_t)}
\\
& &\times 
\frac 
{p^{\bar{\alpha} \theta}_{X_t|U_tY_t}(x_t|u_t,y_t)}
{q^{\bar{\alpha} \theta}_{X_t|U_tY_tZ_t}(x_t|u_t,y_t,z_t)}
\frac
{p^{4\alpha \theta}_{X_t|Y_t}(x_t|y_t)}
{q^{4\alpha \theta}_{X_t|Y_tU_t}(x_t|u_t,y_t)}
{\rm e}^{4\alpha \mu \theta d(x_t,z_t)}
\eeqno
such that they are the distributions 
induced by $q_{U_tX_tY_tZ_t}$. Then for each $t=1,2,$ $\cdots, n$, 
we have the following chain of inequalities:
\beqa
& & \Lambda_{t,q^t}^{(\alpha,\mu,\theta)}
\nonumber\\
&=&
{\rm E}_{q_t}
\left[
\left\{
\frac 
{p_{X_t}^{\bar{\alpha}\theta}(X_t)}
{q_{X_t}^{\bar{\alpha}\theta}(X_t)}
\frac 
{p^{\bar{\alpha} \theta}_{Y_t|X_t}(Y_t|X_t)}
{q^{\bar{\alpha} \theta}_{Y_t|X_tU_t}(Y_t|X_t,U_t)}
\right.\right\}
\nonumber\\
& &\times \left\{
\frac 
{p^{\bar{\alpha} \theta}_{X_t|U_tY_t}(X_t|U_t,Y_t)}
{q^{\bar{\alpha} \theta}_{X_t|U_tY_tZ_t}(X_t|U_t,Y_t,Z_t)}
\right\}
\nonumber\\
& &\times \left\{
\left.
\frac
{p^{4\alpha{\bar{\mu}}\theta}_{X_t|Y_t}(X_t|Y_t)}
{q^{4\alpha{\bar{\mu}}\theta}_{X_t|Y_tU_t}(X_t|U_t,Y_t)}
{\rm e}^{4\alpha \mu \theta d(X_t,Z_t)}\right\}
\right]
\nonumber\\
&=&
{\rm E}_{q_t}
\left[
\left\{
\frac 
{p_{X_t}^{\bar{\alpha}}(X_t)}
{q_{X_t}^{\bar{\alpha}}(X_t)}
\frac 
{p^{\bar{\alpha} }_{Y_t|X_t}(Y_t|X_t)}
{q^{\bar{\alpha} }_{Y_t|X_tU_t}(Y_t|X_t,U_t)}
\right.\right\}^{\theta}
\nonumber\\
& &\times \left\{
\frac 
{q^{\bar{\alpha} }_{X_t|U_tY_t}(X_t|U_t,Y_t)}
{q^{\bar{\alpha} }_{X_t|U_tY_tZ_t}(X_t|U_t,Y_t,Z_t)}
\right\}^\theta
\nonumber\\
& &\times \left\{
\frac 
{p^{4\alpha{\bar{\mu}}}_{X_t|Y_t}(X_t|Y_t)}
{q^{4\alpha{\bar{\mu}}}_{X_t|Y_tU_t}(X_t|U_t,Y_t)}
{\rm e}^{4\alpha \mu d(X_t,Z_t)}\right\}^\theta
\nonumber\\
& &
\times \left.\left\{
\frac
{p_{X_t|U_tY_t}(X_t|U_t,Y_t)}
{q_{X_t|U_tY_t}(X_t|U_t,Y_t)}
\right\}^{\bar{\alpha} \theta}
\right]
\nonumber\\
&\MLeq{a}&
\left(
{\rm E}_{q_t}
\left[
\left\{
\frac 
{p_{X_t}^{\ByShun{\bar{\alpha}}}(X_t)}
{q_{X_t}^{\ByShun{\bar{\alpha}}}(X_t)}
\frac 
{p^{\ByShun{\bar{\alpha}} }_{Y_t|X_t}(Y_t|X_t)}
{q^{\ByShun{\bar{\alpha}} }_{Y_t|X_tU_t}(Y_t|X_t,U_t)}
\right\}^{\frac{\theta}{1-\bar{\alpha}\theta}}
\right.
\right.
\nonumber\\
& &\times \left\{
\frac 
{q^{\bar{\alpha} }_{X_t|U_tY_t}(X_t|U_t,Y_t)}
{q^{\bar{\alpha} }_{X_t|U_tY_tZ_t}(X_t|U_t,Y_t,Z_t)}
\right\}^{\frac{\theta}{1-\bar{\alpha}\theta}}
\nonumber\\
& &\left.\left. \times \left\{
\frac
{p^{4\alpha{\bar{\mu}}}_{X_t|Y_t}(X_t|Y_t)}
{q^{4\alpha{\bar{\mu}}}_{X_t|Y_tU_t}(X_t|U_t,Y_t)}
{\rm e}^{{4\alpha}\mu d(X_t,Z_t)}\right\}^{\frac{\theta}{1-\bar{\alpha}\theta}}
\right]\right)^{1-\bar{\alpha}\theta}
\nonumber\\
& &\times \left(
{\rm E}_{q_t}
\left\{
\frac
{p_{X_t|U_tY_t}(X_t|U_t,Y_t)}
{q_{X_t|U_tY_t}(X_t|U_t,Y_t)}
\right\}\right)^{\bar{\alpha} \theta}
\nonumber\\
&=&
\exp\left\{-(1-\bar{\alpha}\theta)
   \Omega^{(\alpha,\mu,\frac{\theta}{1-\bar{\alpha}\theta})}
   (q_t|p_{XY})
\right\}
\nonumber\\
&\ByShun{\MEq{b}}&
\exp\left\{-\frac{1}{1+\bar{\alpha}\lambda}
   \Omega^{(\alpha,\mu,\lambda)}(q_t|p_{XY})
\right\}
\nonumber\\
&\MLeq{c}&
\exp\left\{-\frac{1}{1+\bar{\alpha}\lambda}
   \hat{\Omega}_n^{(\alpha,\mu,\lambda)}(p_{XY})
\right\}
\nonumber\\
&\MEq{d}&
\exp\left\{-
\frac{1}{1+\bar{\alpha}\lambda}
   {\Omega}^{(\alpha,\mu,\lambda)}(p_{XY})
\right\}.
\label{eqn:sssto} 
\eeqa
Step (a) follows from H\"older's inequality. 
Step (b) follows from (\ref{eqn:Zqdd}).
Step (c) follows from 
the definition of 
$\hat{\Omega}_n^{(\alpha,\mu,\lambda)}$ $(p_{XY})$. 
Step (d) follows from that by Property \ref{pr:pro1} 
part a), the bound $|{\cal U}|\leq |{\cal X}|$$|{\cal Y}|$$|{\cal Z}| $
is sufficient to describe 
$\hat{\Omega}_n^{(\alpha, \mu,\lambda)}($ $p_{XY})$.
Hence, we have the following:
\beqa
& &\max_{q^n\in {\cal Q}^n}\frac{1}{n}
\Omega^{(\alpha,\mu, \theta)}_{\pNqN}
{\ARgRv}
\nonumber\\
&\geq&\frac{1}{n}\Omega^{(\alpha,\mu, \theta)}_{\pNqN}
{\ARgRv}
\MEq{a}-\frac{1}{n}\sum_{t=1}^n \log 
\Lambda_{t,q^t}^{(\alpha,\mu,\theta)}
\nonumber\\
&\MGeq{b}&\frac{1}{1+\bar{\alpha}\lambda}
{\Omega}^{(\alpha,\mu,\lambda)}(p_{XY}).
\qquad \label{eqn:aQ1}
\eeqa
Step (a) follows from (\ref{eqn:uzapa}) in Lemma \ref{lm:keylm}.
Step (b) follows from (\ref{eqn:sssto}).
Since (\ref{eqn:aQ1}) holds for any ${n\geq 1}$ 
and any $p^{(n)}\in {\cal P}^{(n)}$ $(p_{XY})$, we have  
$$
\underline{\Omega}^{(\alpha, \mu,\theta)}(p_{XY})
\geq 
\frac{1}{1+\bar{\alpha}\lambda}{\Omega}^{(\alpha,\mu,\lambda)}(p_{XY}).
$$
Thus we have (\ref{eqn:zzar}) in Proposition \ref{pro:mainpro}.
\hfill\IEEEQED

{\it Proof of Theorem \ref{Th:main}: }
For $\theta \in (0,1/\bar{\alpha})$, set 
\beq
\lambda=\frac{\theta}{1-\bar{\alpha}\theta} 
\Leftrightarrow \theta=\frac{\lambda}{1+\bar{\alpha} \lambda}. 
\label{eqn:abadd}
\eeq
Then we have the following:
\beqno
& &
\frac{1}{n}\log\left\{
\frac{5}{{\rm P}_{\rm c}^{(n)}(\varphi^{(n)},\psi^{(n)};{\Dist})}
\right\}
\\
&\MGeq{a}& 
\frac{
\underline{\Omega}^{(\alpha,\mu,\theta)}(p_{XY})
-4\alpha\theta(\bar{\mu}R+ \mu{\Dist})
}
{1+\theta(3+\alpha-4\alpha\mu)}
\\
&\MGeq{b}& 
\frac{
\frac{1}{1+\bar{\alpha}\lambda}
{\Omega}^{(\alpha,\mu,\lambda)}(p_{XY})
-\frac{4\alpha\lambda}{1+\bar{\alpha}\lambda}(\bar{\mu}R+ \mu{\Dist})
}
{1+\frac{\lambda}{1+\bar{\alpha}\lambda}(3+\alpha-4\alpha\mu)}
\\
&=&
\frac{\Omega^{(\alpha,\mu,\lambda)}(p_{XY})
-4\alpha\lambda(\bar{\mu}R+ \mu{\Dist})}
{1+\bar{\alpha}\lambda+\lambda(3+\alpha-4\alpha\mu)}
\\
&=&F^{(\alpha,\mu,\lambda)}(\bar{\mu}R+\mu{\Dist}|p_{XY}). 
\eeqno
Step (a) follows from Corollary \ref{co:corOne}. Step (b) follows from 
Proposition \ref{pro:mainpro} and (\ref{eqn:abadd}). 
Since the above bound holds for any positive 
$\alpha,$ $\mu,$ and $\lambda$, 
we have 
$$ 
\frac{1}{n}\log\left\{
\frac{5}{{\rm P}_{\rm c}^{(n)}(\varphi^{(n)},\psi^{(n)};{\Dist})}
\right\}
\geq F(R,\Dist|p_{XY}). 
$$
Thus (\ref{eqn:mainIeq}) in Theorem \ref{Th:main} is proved. 
\hfill\IEEEQED

\ProofCor

\section*{\empty}
\appendix

\ApdaAaaa

\ApdaAAA
\ApdaAAAb
\ApdaAAB
\ApdaAABz
\ApdaAAC

\Apda
\Apdb
\Apdc
\vspace*{3mm}

\noindent
\section*{Acknowledgement}

I am very grateful to Dr. Shun Watanabe for his 
helpful comments.

\end{document}